\documentclass[lettersize,journal]{IEEEtran}
\usepackage{amsmath,amsfonts}
\usepackage{algorithm}
\usepackage{array}
\usepackage[caption=false,font=normalsize,labelfont=sf,textfont=sf]{subfig}
\usepackage{textcomp}
\usepackage{stfloats}
\usepackage{url}
\usepackage{verbatim}
\usepackage{graphicx}
\usepackage{cite}
\usepackage{booktabs}
\usepackage{mdframed}
\usepackage{xcolor}
\usepackage{multirow}
\usepackage{hyperref}

\usepackage{fancyvrb}         % 增强的代码排版
\usepackage[noend]{algpseudocode}  % 提供算法伪代码环境，支持灵活的算法描述

\newcommand{\Name}{Prochemy}

\newcounter{finding}
\newcommand{\finding}[1]{\refstepcounter{finding}
  \vspace{2mm}
 \begin{mdframed}[linecolor=gray,roundcorner=12pt,backgroundcolor=gray!15,linewidth=3pt,innerleftmargin=2pt, leftmargin=0cm,rightmargin=0cm,topline=false,bottomline=false,rightline = false]
  #1
 \end{mdframed}
 \vspace{2mm}
}

\usepackage{xspace}

\begin{document}

\title{Prompt Alchemy: Automatic Prompt Refinement for Enhancing Code Generation}


\author{
    \IEEEauthorblockN{
        Sixiang Ye\IEEEauthorrefmark{1},
        Zeyu Sun\IEEEauthorrefmark{2},
        Guoqing Wang\IEEEauthorrefmark{3},
        Liwei Guo\IEEEauthorrefmark{1},
        Qingyuan Liang\IEEEauthorrefmark{3},
        Zheng Li\IEEEauthorrefmark{1},
        Yong Liu\IEEEauthorrefmark{1}
    }
    
    \IEEEauthorblockA{
        \IEEEauthorrefmark{1}Beijing University of Chemical Technology \\
        Beijing, China \\
        Emails: yesx.sxye@gmail.com, liwei.glw@outlook.com, lizheng@mail.buct.edu.cn, lyong@mail.buct.edu.cn
    }
    
    \IEEEauthorblockA{
        \IEEEauthorrefmark{2}National Key Laboratory of Space Integrated Information System \\
        Institute of Software, Chinese Academy of Sciences \\
        Beijing, China \\
        Email: zeyu.zys@gmail.com
    }
    
    \IEEEauthorblockA{
        \IEEEauthorrefmark{3}Peking University \\
        Beijing, China \\
        Emails: guoqingwang@stu.pku.edu.cn, liangqy@stu.pku.edu.cn
    }
        % <-this % stops a space
\thanks{Manuscript received April 19, 2021; revised August 16, 2021.}}

\markboth{IEEE TRANSACTIONS ON SOFTWARE ENGINEERING,~Vol.~14, No.~8, August~2021}%
{Shell \MakeLowercase{\textit{et al.}}: A Sample Article Using IEEEtran.cls for IEEE Journals}


\maketitle

\begin{abstract}
Code generation has gained increasing attention as a task to automate software development by transforming high-level descriptions into executable code. While large language models (LLMs) are effective in generating code, their performance heavily relies on the quality of input prompts. Current prompt engineering methods involve manual effort in designing prompts, which can be time-consuming and yield inconsistent results, potentially constraining the efficacy of LLMs in practical applications. This paper introduces \Name{}, a novel approach for automatically refining prompts iteratively to enhance code generation. \Name{} addresses the limitations of manual prompt engineering by automating the optimization process, ensuring prompt consistency during inference, and aligning with multi-agent systems. It iteratively refines prompts based on model performance, using an optimized final prompt to improve consistency and reliability across tasks. 
We evaluate \Name{} on both natural language-based code generation and code translation tasks using three series of LLMs. Results show that when combining \Name{} with existing approaches, it outperforms baseline prompting methods. It achieves improvements of 5.0\% (GPT-3.5-Turbo) and 1.9\% (GPT-4o) over zero-shot baselines on HumanEval. For the state-of-the-art LDB, \Name{} + LDB  outperforms standalone methods by 1.2–1.8\%. For code translation, \Name{} elevates GPT-4o’s performance on Java-to-Python (AVATAR) from 74.5 to 84.1 (+12.9\%) and Python-to-Java from 66.8 to 78.2 (+17.1\%). Furthermore, considering that the o1-mini model integrates prompt engineering techniques, \Name{} can continue to show good performance among it, further validating its effectiveness in code generation and translation tasks. 
Additionally, \Name{} is designed to be plug-and-play, optimizing prompts with minimal human intervention and seamlessly bridging the gap between simple prompts and complex frameworks.

\end{abstract}


\section{Introduction}

\IEEEPARstart{C}{ode} generation has become a transformative application in software engineering, allowing the automatic generation of code from high-level descriptions or specifications~\cite{chen2021evaluating, svyatkovskiy2020intellicode, 10.1145/3690635}. This capability accelerates the development process and reduces human error by automating repetitive coding tasks. As the demand for efficient software development grows, the use of large language models (LLM) for code generation has become increasingly significant~\cite{fan2023largelanguagemodelssoftware}. 

To effectively use LLM, prompt engineering plays an important role, involving design and refinement of input prompts to optimize model performance and improve the quality of generated code~\cite{liu2024logprompt, fu2023chainofthoughthubcontinuouseffort}. Early prompt engineering approaches focus on natural language processing tasks, including text generation and sentiment analysis~\cite{guo2024connectinglargelanguagemodels, wu2024infoprompt}. In these tasks, few-shot learning enables models to adapt to new tasks with minimal examples~\cite{radford2019language, mann2020language}. The Chain of Thought (CoT) method enhances precision in complex tasks by incorporating reasoning steps into prompts~\cite{fu2023chainofthoughthubcontinuouseffort}. 
Recently, researchers who supplement prompts with additional code contextual information~\cite{10.1145/3690635} or adopting multi-agent system approaches to debugging and regenerating existing code~\cite{huang2024agentcoder, zhong2024debuglikehumanlarge}, these methods demonstrate improvements in the quality of code results.

Although prompt-based techniques have shown their effectiveness, their limitations become apparent in the realm of code generation. Such tasks often involve intricate contextual demands, and different coding challenges may hinge on specific aspects of the prompt. 

However, existing approaches depend on manually crafting optimal prompts, which is time consuming and susceptible to errors. Human intuition often fails to meet the stringent demands of coding standards, which could lead to suboptimal results. Furthermore, human-designed prompts may not align consistently with the specific needs of LLM, which can compromise performance~\cite{10.1145/3597503.3639183}. 

To mitigate these problems and enhance code generation, the necessity for an automated prompt optimization approach becomes evident. This approach must address two key challenges. 
The first challenge is \textbf{automation}: it should leverage algorithms to analyze coding contexts, automatically generating prompts that accurately capture the specific needs of LLMs on diverse coding tasks. Once generated, the prompt must be fixed for use during inference, ensuring that it can be reused without adjustments. %avoiding any adjustments at that stage to ensure consistency and reliability in predictions.
The second challenge is \textbf{compatibility}: the approach should ensure alignment with existing techniques, such as multi-agent approaches. This compatibility will enable integration into current workflows, ultimately enhancing both performance and efficiency in code generation.

%% 合并
%% prochemy使用执行信息来实现refine过程
% \Name{} begins with an initial prompt, such as a zero-shot prompt, and iteratively refines it based on the model’s performance. This refinement uses a training set derived from existing datasets and their generated variations. The process aims to align the prompt with the needs of LLMs, enhancing its accuracy for specific tasks. After these refinements, the prompt is finalized and set, ensuring it can be reused consistently without further adjustments.

To address these challenges, we propose \Name{} (derived from \underline{Pro}mpt Al\underline{chemy}), a novel execution-driven automated prompt generation framework designed to guide LLMs in code generation tasks. For \textbf{automation}, \Name{} begins with an initial prompt, such as a zero-shot prompt, and iteratively refines it based on the model’s performance evaluated by execution. This refinement uses a training set derived from existing datasets and their generated variations. The process aims to align the prompt with the needs of LLMs, enhancing its accuracy for specific tasks. After these refinements, the prompt is finalized and set, ensuring it can be reused consistently without further adjustments.
For \textbf{compatibility}, \Name{} is designed as a plug-and-play approach that integrates seamlessly without requiring any modifications to existing prompt methodologies, such as CoT and multi-agent systems.

We conduct experiments on natural-language-based code generation and code translation tasks using three series of LLMs: the ChatGPT series~\cite{radford2018improving, radford2019language, brown2020language}, the Claude series~\cite{bai2022constitutionalaiharmlessnessai, bai2022traininghelpfulharmlessassistant}, and the DeepSeek series~\cite{deepseekai2024deepseekv2strongeconomicalefficient, guo2024deepseekcoderlargelanguagemodel, deepseekai2024deepseekv3technicalreport}. \Name{} enhances code generation and translation across LLMs, achieving average gains of $+4.04\%$ (zero-shot), $+4.55\%$ (chain-of-thought), and $+2.00\%$ (multi-turn) on HumanEval/MBPP benchmarks. Correctness-critical variants (HumanEval+/MBPP+) see $+4.21–7.76\%$ improvements, with LDB integration reaching state-of-the-art results (96.3\% on HumanEval). On LiveCodeBench, \Name{} delivers $+14.15\%$ (zero-shot) and $+12.28\%$ (chain-of-thought) average gains. For code translation, it achieves $+13.68\%$ (Java-Python) and $+8.25\%$ (Python-Java) average improvements on CodeNet and AVATAR datasets. Considering that the o1-mini model integrates CoT techniques, we subsequently applied the same \Name{} approach to this model. \Name{} continues to show good performance, further validating its effectiveness in code generation and translation tasks.
These results validate \Name{}’s unified framework, which aligns prompts with task requirements and model capabilities without architectural changes, establishing a scalable paradigm for LLM-driven code tasks.

We summarize our contributions in this paper as follows.
% \vspace{-3mm}
\begin{itemize}
    \item We propose \Name{}, the first approach designed to specifically optimize prompts for code generation tasks. It iteratively refines prompts by analyzing the model’s performance against a training set composed of existing datasets and their generated variations. \Name{} requires only a single training run for a given task. After the optimization process, the prompt is fixed as the final prompt.
    \item \Name{} is compatible with existing prompt engineering methodologies, offering a plug-and-play solution that integrates seamlessly into current frameworks. This adaptability allows for its use without requiring modifications to established workflows.
    \item We conduct extensive empirical validation of \Name{} across multiple datasets and language models, demonstrating consistent improvements in performance for both natural-language-based code generation and code translation tasks. This evaluation confirms the robustness and generalizability of \Name{} across diverse architectural frameworks.
\end{itemize}

\section{Related Work}
\label{sec:relatedwork}
Fine-tuning large language models (LLMs) for different domains is a daunting task~\cite{10.1145/3520312.3534862}, particularly when dealing with the sheer volume of parameters, such as DeepSeek-V$3$, which contains $671$ billion parameters~\cite{deepseekai2024deepseekv3technicalreport}. In response, prompting techniques have emerged as a preferred strategy to adapt LLMs for specific tasks by supplying a specialized prompt~\cite{10.1145/3540250.3549113}.

%% zero-shot and few-shot
\textbf{Prompt Engineering for Natural Language. } Early efforts in prompt engineering for LLMs revolved around zero-shot and few-shot prompting~\cite{radford2019language, mann2020language}. 
% In zero-shot prompting, the task description, generating a response without any prior examples.
In zero-shot prompting, the model is required to generate code directly based on the task description, without any prior examples.
This method enables LLMs to tackle diverse tasks but may lack accuracy in more complex scenarios. Few-shot prompting extends this technique by incorporating a few example pairs of tasks and corresponding responses, allowing the model to learn from these samples and adjust its outputs accordingly. This example-based approach enhances the model's accuracy, leveraging the provided information to guide its reasoning process.

%% CoT, APE, APO and OPRO
Researchers have advanced prompt engineering to enhance models' reasoning capabilities. Chain-of-Thought (CoT) prompting~\cite{wei2022chain} instructs models to articulate intermediate reasoning steps, improving performance on tasks that require logical, step-by-step problem-solving. This technique ensures more coherent and accurate handling of complex tasks.

Automated methods like the Automatic Prompt Engineer (APE) automate the optimization of LLM instructions, often exceeding the performance of human-written prompts by refining model behavior to enhance results like informativeness and truthfulness~\cite{zhou2023largelanguagemodelshumanlevel}. Similarly, ProTeGi uses gradient descent methodology to optimize prompts automatically, improving task performance by up to 31\%~\cite{pryzant2023automaticpromptoptimizationgradient}. The LLM as the Optimizer (OPRO) method uses LLMs' natural language understanding to iteratively refine prompts and enhance solution quality. Without relying on gradient-based techniques, it adjusts solutions based on past evaluations, effectively balancing exploration and exploitation to improve problem-solving efficiency~\cite{yang2024largelanguagemodelsoptimizers}.

Black-Box Prompt Optimization (BPO)~\cite{cheng2024blackboxpromptoptimizationaligning} enhances LLM alignment for the natural language processing tasks by optimizing user prompts based on human preferences without modifying model parameters. It boosts model output quality across various tasks and outperforms methods like PPO~\cite{schulman2017proximalpolicyoptimizationalgorithms} and DPO~\cite{rafailov2024directpreferenceoptimizationlanguage}, offering an efficient and interpretable alignment solution.

\textbf{Prompt Engineering for Code Generation. }
In the domain of software engineering, traditional prompt engineering methods designed for natural language tasks face limitations due to the need for strict adherence to standards~\cite{fan2023largelanguagemodelssoftware}.

%% SCoT, AgentCoder and LDB
To overcome these challenges, researchers have developed specialized approaches that integrate structured reasoning and multi-agent systems. For example, SCoT (Structured Chain-of-Thought) leverages program architecture to enhance code generation accuracy by using programmatic structures~\cite{10.1145/3690635}, while agent-based systems like AgentCoder combine specialized agents—such as test designers, executors, and programmers—to automate testing and improve code quality~\cite{huang2024agentcoder}. Self-collaboration frameworks also simulate human software development processes, optimizing models for complex programming tasks~\cite{dong2024selfcollaborationcodegenerationchatgpt}. Furthermore, techniques like LDB (Debug-Like-Human) use runtime information and program segmentation to improve debugging~\cite{zhong2024debuglikehumanlarge}. Despite their effectiveness, these methods often struggle with high token consumption, high iteration counts, and limited applicability across different models.

%% how do our method different from these methods
Our approach is designed to create a flexible prompt optimization pipeline that is suitable for a range of models. This pipeline refines prompts by utilizing insights gained from program execution, aiming to discover the most effective discrete prompt for code generation. This approach can be integrated with existing prompt engineering techniques to achieve improved results.

\section{Approach}
\label{sec:approach}
% \vspace{-1mm}

\begin{figure*}[t]
\centering
\includegraphics[width=.82\textwidth]{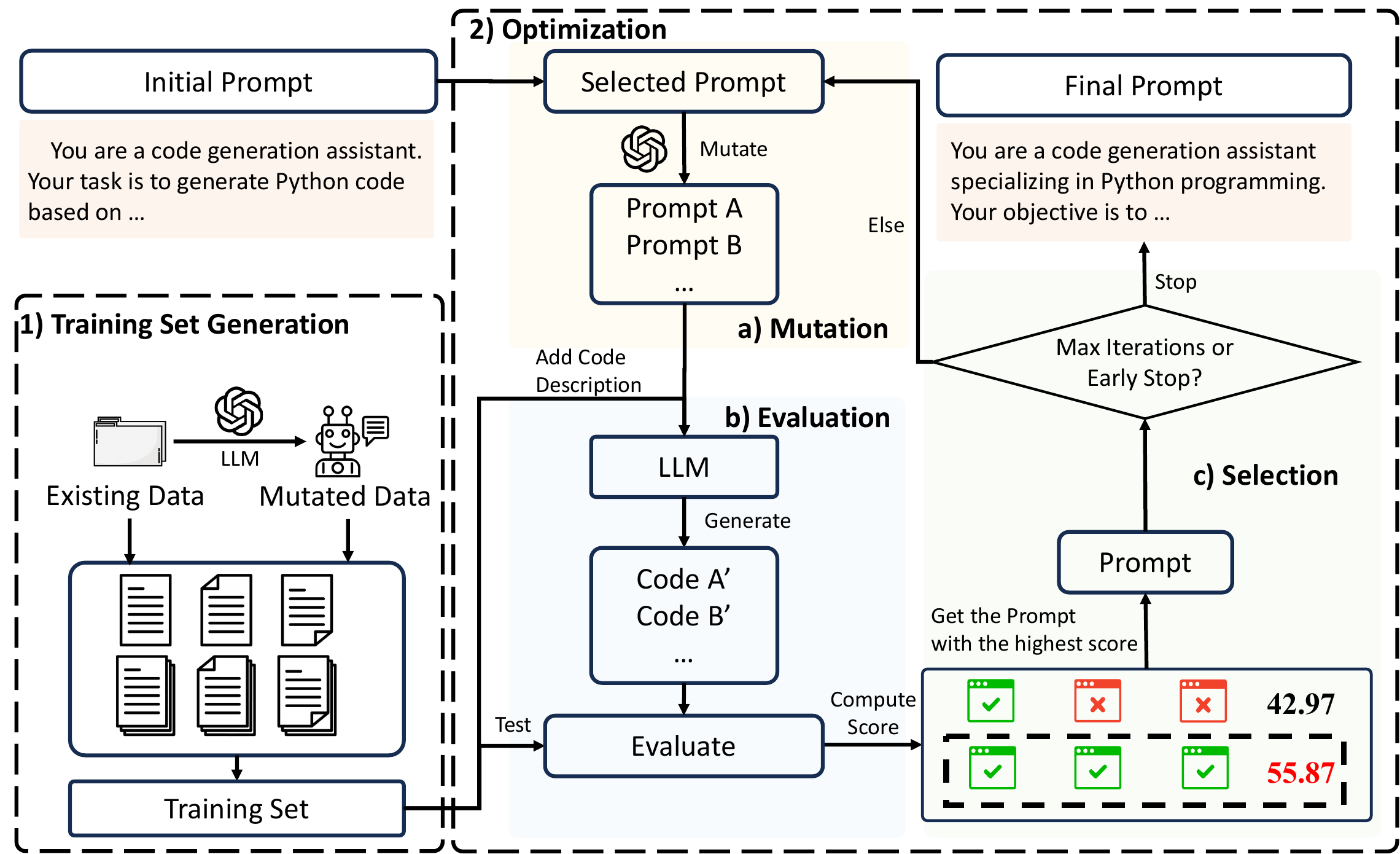}
% \vspace{-0.4cm}
\caption{Overview of \Name{}}
% \vspace{-0.6cm}
\label{Overview}
\end{figure*}

\subsection{Overview}
%% 图需要重构一下，体现出方法是迭代的
To optimize prompts in a plug-and-play manner and iteratively refine them based on the model's performance, \Name{} involves two steps, as illustrated in Fig.~\ref{Overview}:
1) \textbf{Training Set Generation} aims to generate the training set for evaluating the effectiveness of prompts (in Sec.~\ref{train_set_generation}).
2) \textbf{Optimization} aims to iteratively refine prompts through three sub-steps~\ref{Optimization}:
   a) \textbf{Mutation}—Generates variations of existing prompts to introduce diversity and explore different structures, providing a pool of candidates for further evaluation (in Sec.~\ref{prompt_mutation}).
   b) \textbf{Evaluation}—Assesses the performance of these mutated prompts using the training data, employing a weighted averaging mechanism to determine their effectiveness (in Sec.~\ref{prompt_evaluation}).
   c) \textbf{Selection}—Chooses the most effective prompt from the pool based on their evaluated performance, guiding the subsequent mutation process (in Sec.~\ref{prompt_selection}).

In particular, \Name{} requires only a single training run for a given task. After the optimization process, the prompt is finalized and set as the final prompt. This streamlined approach not only enhances efficiency but also ensures that the refined prompt is tailored for optimal performance and can be reused consistently without further adjustments in the context of the task at hand.

% (2) \textbf{Prompt Mutation} generates variations of existing prompts to introduce diversity and explore different prompt structures, setting up a pool of candidates for evaluation in the next step (in Sec.~\ref{prompt_mutation}).
% (3) \textbf{Prompt Evaluation} assesses the performance of the mutated prompts using training data, applying a weighted averaging mechanism (in Sec.~\ref{prompt_evaluation}).
% (4) \textbf{Prompt Selection} aims to select the most effective prompt in each iteration, providing a reference for the subsequent mutation process (in Sec.~\ref{prompt_selection}).

\subsection{Training Set Generation}
\label{train_set_generation}
%% 生成的数据并非都是正确的，这样的情况怎样处理
%% data的mutate是怎么做的
%% 这一步的作用，和后续步骤的联系
The training set generation step is designed to create a dataset $T$ that facilitates the indirect evaluation of prompt quality. 
In the context of code generation, this step constructs the dataset where each data entry $T_i \in T$ consists of an input natural language description $T_i^{(NL)}$ and its corresponding set of executable test cases $T_i^{(test)}$, ensuring that each description is actionable and testable ($T_i = (T_i^{(NL)}, T_i^{(test)})$). 
This dataset is derived from two primary sources: existing training data from datasets, and new data generated by LLMs through mutation processes applied to pre-existing data. 

\paragraph{Existing Data} For existing data, we construct the training set \( D_{\text{existing}} \) via implementing a sampling strategy. This involves randomly selecting samples from existing datasets until $K$ samples are obtained\footnote{In our experiments, the selected datasets do not overlap with our test dataset (\( D_{\text{test}} \)), ensuring the independence of our training and test datasets.}.

%% 补充了中间流程
\paragraph{Mutated Data} To enhance the diversity of our training set, we utilize an automated data augmentation mechanism that generates new data through mutations by LLMs. \Name{} employs LLMs to create additional samples by mutating reference data, which is randomly sampled from existing datasets but does not overlap with the previously mentioned existing data \( D_{\text{existing}} \). Specifically, the initial input for mutation is a random sample from the existing dataset, ensuring that the mutated data is different from the existing ones.

%% 怎么得到dgen的：对特定的参考数据进行重复变异，将每一轮的结果保留作为dgen的内容，这个过程不涉及到选择，结果都会保留
Formally, we sample \(L\) distinct reference data from the existing datasets, ensuring no overlap with \(D_{\text{existing}}\). For each reference data, we use the LLM with a specific prompt to mutate it into new samples that are different from the existing data. The specific prompt is as follows:
\textit{"You are an expert in software engineering. Please help me generate similar data based on the format provided below. The reference data format is as follows."}
This process is repeated \(K\) times, resulting in a total of \(K\) generated samples. To ensure correctness, each generated sample undergoes validation through two steps: (1) executing the code to verify functional integrity, and (2) evaluating against model-provided test cases. Only samples passing both validation phases are retained. In this paper, we set \(K = 10\). The resulting samples are retained as the content of \(D_{\text{gen}}\).

% It is important to note that the data generated through mutation may not be entirely correct, and some generated samples may fail to meet the task requirements or pass validation tests. However, since all prompts are evaluated on the same training set \( T \), which includes both accurate and inaccurate mutated data, any potential errors in the generated data affect all prompts equally. Therefore, these inaccuracies do not bias the subsequent prompt evaluation process and allow for a fair comparison of prompt performance.

The final training set is a combination of these two data sources, denoted as \( T = D_{\text{existing}} \cup D_{\text{gen}} \). This training set \( T \) serves as the foundation for further evaluation of prompt quality and performance.

\subsection{Optimization}
\label{Optimization}
The optimization step of \Name{} focuses on refining prompts iteratively to maximize their effectiveness for code generation tasks. This phase involves three sub-steps, each designed to progressively enhance the quality of prompts through mutation, evaluation, and selection. These iterative cycles are aimed at continuously improving prompt performance, adapting to evolving task requirements, and learning from previous iterations. The process is designed to repeat for a maximum of \( k_\text{max} \) iterations, ensuring thorough exploration while preventing excessive computation. In this paper, \( k_\text{max} \) is set to 10.

\subsubsection{Mutation}
\label{prompt_mutation}
%% 生成很多个prompt，怎样选取其中的M个
The primary goal of the mutation process is to generate a diverse set of prompts by mutating existing ones. This approach facilitates the creation of a broad range of prompt variations, allowing for a comprehensive exploration and subsequent selection of the most effective prompts to enhance task performance.

%% 结束条件是什么，N轮或者收敛，输出XXX作为结果
%% Prompt Mutation的目的及方法概述
%% 对于第一次迭代，怎么做
This process is achieved by generating a series of modified prompt variants \( P_i^{(k)} \) based on prompts selected from the previous iteration. In particular, for the first iteration, the prompt can be set to the simple zero-shot prompt or the well-designed prompt of existing approaches. Each variant \( P_i^{(k)} \) retains the core semantic structure but incorporates changes in linguistic or structural elements, allowing for a detailed evaluation of how these variations influence the performance of the LLM. The modifications include adjustments such as rephrasing sentences, refining clarity, or altering the presentation of task instructions.

%% 解释了一下第一次时集合s是什么情况
Formally, given the set of selected prompts $S^{(k-1)}$  from iteration \( k - 1 \) (for the first iteration, $S^{(0)}$ contains only an initial prompt like zero-shot or CoT), the mutation process produces a new set $P^{(k)}$ ($|P^{(k)}| = n$ ) of prompt variants for iteration \( k \):

\[
P^{(k)} = \{ P_1^{(k)}, P_2^{(k)}, \ldots, P_n^{(k)} \},
\]

where each \( P_i^{(k)} \) is generated by applying a mutation function \( m \) to a randomly sampled prompt \( S_j^{(k-1)} \) from the selected set \( S^{(k-1)} \):

\[
P_i^{(k)} = m(S_j^{(k-1)}), \quad S_j^{(k-1)} \in S^{(k-1)}.
\]

%% 解释了一下转换函数m的具体工作方式
The transformation function \( m(\cdot) \) introduces variations into the prompts by adding modifications to the original prompt $S_j^{(k-1)}$. These changes are designed to explore different expressions while preserving the original intent of the prompt. The specific prompt is as follows: \textit{You are an expert prompt engineer. Please help me improve the given prompt to get a more helpful and harmless response.}

\subsubsection{Evaluation}
\label{prompt_evaluation}
%% 使用了执行信息
The evaluation process systematically assesses the effectiveness of mutated prompts by analyzing execution results of their generated code implementations, including metrics of functional correctness and runtime performance. This execution-based analysis enables the identification of optimal prompt variants in each iteration through quantitative comparisons of code behavior across different problem instances.

%% C是针对当前任务最常用的评价指标
To achieve this, \Name{} applies each mutated prompt \(P_i^{(k)}\) to every dataset entry \(T_j = (T_i^{(NL)}, T_i^{(test)})\) in the generated training set \(T\). It concatenates the prompt \(P_i^{(k)}\) with the natural language description \(T_i^{(NL)}\) of each task. Then, \Name{} feeds this concatenated input into the LLM, which produces outputs \(R_{ij}^{(k)}\). Each output \(R_{ij}^{(k)}\) is evaluated against the associated test suite \(T_i^{(test)}\) using specific criteria \(C\), resulting in a score \(M_{ij}\) for each prompt-task pair. This process ensures that the effectiveness of each prompt is quantitatively assessed in the context of its ability to generate correct and functional code. Here, \( C \) denotes execution-based evaluation metrics derived from code runtime verification. In code generation tasks, we employ the pass@1 metric where \( M_{ij} \in \{0, 1\} \) is determined through automated test case execution - assigning 1 only if the generated code passes all ground-truth test cases on initial execution. This execution-driven paradigm ensures metrics fundamentally reflect functional correctness rather than surface-level code similarity.

%\Name{} applies each mutated prompt \(P_i^{(k)}\) to every data $T_j  = (T_i^{(NL)}, T_i^{(test)})$ in the generated training set \(T\). \Name{} concatenates the prompt $P_i^{(k)}$ with $T_i^{(NL)}$. Then, \Name{} feeds the concatenated prompt into the LLM, producing outputs \(R_{ij}^{(k)}\). Each output \(R_{ij}^{(k)}\) is then evaluated based on the test suite $T_i^{(test)}$ against specific criteria \(C\) to determine its quality, resulting in a score \(M_{ij}\) for each prompt-task pair. 

\paragraph{Weighted Scoring Mechanism.} To optimize the evaluation process and facilitate informed prompt selection, a weighted scoring mechanism is employed. This mechanism prioritizes the evaluation of each dataset entry \(T_j\) based on its complexity. The underlying principle is that if the code generated by a multitude of prompts correctly solves a particular dataset entry \(T_j\), this suggests that the data may be relatively simple and less critical. Consequently, less weight is assigned to such data entries, reflecting their lower importance in the overall assessment of prompt effectiveness. This approach ensures that more complex and challenging tasks, which are harder to solve correctly, have a greater influence on the selection of the most effective prompts.

Formally, this mechanism assigns weights \( w_j \) to tasks \( T_j \) based on their complexity and the number of times they are successfully solved in this iteration:

\[
w_j = \frac{|P^{(k)}|}{N_{\text{successful}}(T_j)},
\]

where $|P^{(k)}|$ is the total number of prompts variants, and \( N_{\text{successful}}(T_j) \) is the number of times \( T_j \) is successfully solved in this iteration.

The overall weighted score for each prompt \( P_i^{(k)} \) is calculated as:

\[
W_S(P_i^{(k)}) = \sum_j w_j \cdot M_{ij}.
\]

\subsubsection{Selection}
\label{prompt_selection}

After evaluating the performance of each mutated prompt, the prompt selection step focuses on choosing the most effective prompts to carry forward into subsequent iterations or to serve as the final output prompt.

% \paragraph{Iterative Selection Process. }This process is crucial for refining the prompt optimization process, ensuring that the iterative cycle of mutation, evaluation, and selection converges toward an optimal prompt \( P^* \).

In each iteration \( k \), the selection function is applied to the results of evaluation \( W_S(P) \). Specifically, the prompt(s) with the highest weighted average score is selected as the reference for the next round of mutation:

\[
S^{(k)} = \{ P  \mid P = \arg \max_{P' \in P^{(k)}} W_S(P') \}.
\]

If there is a single prompt with the highest score, it is used directly. If multiple prompts share the highest score, all are retained. The selected prompt(s) \( S^{(k)} \) serve as the basis for generating new mutations in the next iteration.

If, upon reaching the convergence criteria, there are multiple prompts with the same highest weighted average score, we randomly select one of these prompts as the final result.

% \paragraph{Selection Criteria.} The selection function \( s \) is defined as selecting the set \( S_k \) of prompts with the highest weighted score in iteration \( k \). During mutation, one prompt \( P_{\text{ref}}^{(k)} \) is randomly selected from \( S_k \) to serve as the reference: \( P_{\text{ref}}^{(k)} \sim \text{Uniform}(S_k) \). Here, \(\text{Uniform}(S_k)\) means that one element is chosen randomly from \( S_k \), with each element having an equal chance of being selected.

% By adopting this strategy, the selection process ensures that only the top-performing prompts influence the next iteration while also introducing variability when multiple prompts are equally effective.

% \paragraph{Integration into Iterative Process.} The selected prompt(s) \( S_k \) serve as the basis for generating new mutations in the next iteration. The iterative process is updated as follows:

% \[
% P^{(k)} \xrightarrow{\text{Evaluation}} W\_S(\textbf{P}^{(k)}) \xrightarrow{\text{Selection}} S_k = \arg\max_{P \in \textbf{P}^{(k)}} W\_S(P)
% \]

% \[
% P_{\text{ref}}^{(k)} \sim \text{Uniform}(S_k)
% \]

% \[
% \textbf{P}^{(k+1)} = m(P_{\text{ref}}^{(k)})
% \]

% By continually selecting high-performing and diverse prompts, the method ensures that the search space is thoroughly explored, and the most effective prompts are identified and refined over successive iterations.

\paragraph{Convergence Toward Optimal Prompt.} The iterative process continues until one of the following termination conditions is met: (1) Early Stop: 
% the improvement of highest weighted scores between successive iterations falls below the threshold set from the previous iteration,
the highest weighted score  remains unchanged for three consecutive iterations, % the improvement in weighted scores between successive iterations falls below a predefined threshold \( \epsilon \),
or (2) Max Iterations: a maximum number of iterations is reached. This approach ensures that the method efficiently converges toward the optimal prompt \( P^* \) without unnecessary computational effort.

Formally, the convergence condition can be defined as:

\[
\begin{gathered}
(W_S(S_j^{(k)}) = W_S(S_j^{(k-1)}) = W_S(S_j^{(k-2)}) \And k \geq 3) \\
\text{or} \quad k \geq k_{\text{max}}.
\end{gathered}
\]

Here, $S_j^{(k)} \in S^{(k)}$ (the score of any $S_j^{(k)} \in S^{(k)}$ is equal), and \( k \) represents the current iteration number.  
% Convergence is considered achieved when either of the following conditions is met: (1) \( \epsilon \) remains at zero for two consecutive iterations; (2) the number of iterations reaches the maximum limit, i.e., \( k \geq k_{\text{max}} \).
Upon reaching one of these termination conditions, the prompt with the highest weighted score is selected as the final output prompt \( P^* \in S_k\). If there are multiple prompts with equally high scores, one is chosen at random to be \( P^*  \in S_k\).

\section{Experimental Setup}
% \vspace{-1mm}
\label{sec:setup}
In this section, we describe the dataset, large language models, evaluation metrics, and experimental platforms. Our evaluation is designed to answer the following research questions.

\textbf{RQ1.} \textit{How does \Name{} perform compared to other methods?} To answer this question, we conduct experiments on two tasks with seven datasets and compare \Name{} with state-of-the-art approaches.

\textbf{RQ2.} \textit{What effects do each component in the \Name{} have?} To answer RQ2, we analyze \Name{} by removing each component to discern its specific contribution to the overall effectiveness of \Name{}.

\textbf{RQ3.} \textit{In what key aspects and to what extent does \Name{} improve models' code capabilities compared to previous methods?} To answer this question, we illustrate a case study, analyzing specific code tasks to assess improvements in accuracy, efficiency, and generalization. We compare these results with previous methods to highlight \Name{}'s key advantages.

\subsection{Dataset}
We evaluate \Name{} on two distinct types of code generation tasks. The first type is Natural-Language-Based Code Generation, which involves generating executable code from natural language descriptions. This task tests the ability of \Name{} to accurately interpret human language and produce functional programming code. The second type is Code Translation, which requires translating a program from one programming language, which serves as the input description in \Name{}, to another. This task assesses the capability of \Name{} to maintain logical and syntactical correctness across different programming languages.

\subsubsection{Natural-Language-Based Code Generation}
%% 这两个数据集关于数据泄露的讨论

%% For the evaluation of code generation task, we selected the HumanEval and MBPP datasets.

%% add a table
% \textbf{Humaneval. }HumanEval is a widely used dataset for testing code generation models, featuring numerous programming challenges with function signatures, docstrings, and test cases to ensure code meets specific requirements~\cite{chen2021evaluating}.

% The dataset evaluates a model's ability to comprehend problem descriptions, generate correct code, and handle edge cases. Challenges typically involve creating a function body from a given signature and assessing its correctness with multiple test cases. With 164 challenges averaging 9.6 tests each, HumanEval simulates real-world software development scenarios.

% \textbf{MBPP. }The MBPP (Mostly Basic Python Problems) dataset~\cite{austin2021programsynthesislargelanguage} consists of 974 programming challenges in Python, spanning basic to intermediate levels of difficulty and encompassing tasks such as string manipulation and numerical calculations. Each challenge includes a detailed description, illustrative examples, and three test cases, which are intended to assess the performance of code generation models in solving practical programming issues. This dataset is particularly appropriate for evaluating the capabilities of code autocompletion tools and AI-based programming assistants in a scholarly context.

We evaluate the effectiveness of \Name{} using five widely adopted natural-language-based code generation datasets: HumanEval~\cite{chen2021evaluating}, 
MBPP~\cite{austin2021programsynthesislargelanguage}
and its enhanced versions, HumanEval+, and MBPP+~\cite{10.5555/3666122.3667065}. To mitigate data leakage during large language model training, we also use LiveCodeBench~\cite{jain2024livecodebench}, a code generation dataset that is updated over time.

\textbf{HumanEval and HumanEval+.} %These datasets feature programming challenges to test a model's problem-solving abilities. HumanEval includes diverse problems requiring Python solutions, while HumanEval-ET extends it with more comprehensive test cases, increasing the difficulty and suitability for evaluating advanced models.
The HumanEval and HumanEval+ datasets each consist of $164$ programming challenges designed to assess a model's problem-solving capabilities in Python. In HumanEval, each problem includes an average of $7.7$ test cases, focusing on fundamental to intermediate concepts. HumanEval+, on the other hand, extends the original dataset with more extensive testing, providing test cases that exceed the original dataset by an average of $80$ times. This increase in test cases raises the difficulty and rigor of the evaluation, making HumanEval+ particularly suitable for benchmarking advanced models with a higher emphasis on robustness and accuracy across complex scenarios.

% For the training set generation of \Name{}, we select the similar MBPP dataset~\cite{dong2024codescoreevaluatingcodegeneration}.

\textbf{MBPP and MBPP+.} These datasets assess a model's Python proficiency and ability to handle various coding tasks. MBPP evaluates fundamental programming skills, and MBPP+ enhances it by providing an average of over $35$ times more test cases than the original, improving the evaluation of functional correctness under diverse conditions.

\textbf{LiveCodeBench.} This dataset addresses the critical challenge of data contamination in LLM evaluation by exclusively curating time-stamped programming problems collected in real-time from competitive coding platforms and technical interviews. To ensure temporal separation from pre-training data, our experiments utilize the official release-v4 lite subset, comprising $101$ representative tasks systematically sampled from problems released between May $2023$ and September $2024$. Each challenge integrates dynamically generated test suites encompassing edge cases, explicitly designed to emphasize real-world problem-solving patterns rather than synthetic benchmarks. By anchoring evaluations to chronologically novel programming challenges unavailable during model training, LiveCodeBench establishes a rigorous framework for assessing genuine model generalization capabilities while mitigating memorization effects inherent in static benchmark reuse. The lite subset maintains statistical parity with the full $713$-problem corpus through stratified sampling across difficulty levels and problem domains.

% \begin{table}[t]
% \centering
% \caption{Statistics of the Datasets used in Code Generation}
% \resizebox{\textwidth}{!}{
% \begin{tabular}{>{\bfseries}l >{\centering}m{2.5cm} >{\centering}m{2.5cm} >{\centering}m{2.5cm} >{\centering\arraybackslash}m{2.5cm}}
%     \toprule
%     \textbf{Statistics} & \textbf{HumanEval} & \textbf{HumanEval-ET} & \textbf{MBPP} & \textbf{MBPP-ET} \\
%     \midrule
%     \textbf{Language} & Python & Python & Python & Python \\
%     \textbf{Samples} & 164 & 164 & 974 & 974 \\
%     \textbf{Avg. tests per sample} & 7.7 & 108 & 3 & 102 \\
%     \bottomrule
% \end{tabular}
% }
% \label{Statistics}
% \end{table}

\subsubsection{Code Translation}

For code translation, we use the CodeNet and AVATAR  dataset to evaluate prompt performance. 

\textbf{CodeNet.} The CodeNet dataset consists of programs in various languages~\cite{puri2021codenetlargescaleaicode}, each solving specific tasks. Programs are labeled as submissions (potentially incomplete or incorrect) or solutions (accepted, executable, and correct across all test cases). This dataset serves for understanding and improving programming practices and automated code evaluation.

\textbf{AVATAR.} AVATAR is a corpus of $9,515$ programming tasks in Java and Python, with $3,391$ parallel independent functions for training and evaluating program translation tasks~\cite{ahmad2023avatarparallelcorpusjavapython}. It includes problems from platforms like CodeForces, AtCoder, AIZU, CodeJam, GeeksforGeeks, LeetCode, and ProjectEuler, and supports execution-based program translation evaluation. 

Following \cite{Pan_2024_lost}, we adopt rigorously validated subsets from two established code repositories: (1) The CodeNet subset comprises $200$ Python and Java programming tasks, each accompanied by implementations in both languages, and (2) the AVATAR subset contains 250 algorithm design challenges. All implementations are verified to pass their respective test cases through automated evaluation pipelines, ensuring functional correctness prior to experimental usage. On average, the AVATAR subset has $25.02$ test cases per task, while CodeNet has $1$ test cases per task.

% For the training set generation of \Name{}, we select the similar CodeNet dataset~\cite{puri2021codenetlargescaleaicode}, which also includes Python and Java implementations for 200 tasks~\cite{Pan_2024_lost}.

% For our experiments, we select subsets from two datasets described in~\cite{Pan_2024_lost}. The AVATAR subset includes Python and Java implementations for 250 tasks, and the CodeNet subset contains implementations for 200 tasks in both languages. All implementations are pre-verified to pass their test cases. On average, the AVATAR subset has 25.02 test cases per task, while CodeNet has 1 test cases per task.

% \begin{table}[t]
% \centering
% \caption{Statistics of the Datasets used in Code Translation}
% \begin{tabular}{l c c}
%     \toprule
%     \textbf{Statistics} & \textbf{AVATAR} & \textbf{CodeNet} \\
%     \midrule
%     \textbf{Language} & Java, Python & Java, Python \\
%     \textbf{Samples} & 250 & 200 \\
%     \textbf{Avg. tests per sample} & 25.02 & 1.295 \\
%     \bottomrule
% \end{tabular}
% \end{table}

\subsection{Baselines}
In our evaluation, we select a set of baseline methods specifically designed for code generation and evaluation tasks. These baselines represent state-of-the-art approaches in the field, categorized broadly into single-turn and multi-turn (or multi-agent) methods.

\paragraph{Single-Turn Methods} rely on a single prompt-response interaction for generating code solutions. This type of method contains: 1) \textbf{Traditional Methods} include basic approaches like Zero-Shot and Chain-of-Thought (CoT) prompting, which represent standard prompting techniques without iterative optimization or specialized prompt engineering.
2) \textbf{Prompt Optimization} includes advanced prompt optimization approaches. The Automatic Prompt Engineer (APE)~\cite{zhou2023largelanguagemodelshumanlevel} generates candidate instructions via LLMs and selects optimal prompts through iterative scoring. Optimization by PROmpting (OPRO)~\cite{yang2024largelanguagemodelsoptimizers} leverages LLMs as natural language optimizers, iteratively refining solutions based on historical evaluations. Black-box Prompt Optimization (BPO)~\cite{cheng2024blackboxpromptoptimizationaligning} aligns LLMs by optimizing user prompts without parameter updates, using human preferences to enhance intent alignment for untrainable models like GPTs. We adopt them due to the absence of code-specific optimization baselines.
%SCoT~\cite{10.1145/3690635}, 

\paragraph{Multi-Turn Methods} To better encode the code knowledge. These approaches involve iterative or collaborative interactions to enhance code generation performance. 

% \textbf{Multi-Turn Methods} include techniques such as ReAct~\cite{yao2022react}, Reflexion\cite{shinn2024reflexion}, and ToT~\cite{yao2024tree} which focus on iteratively refining responses through multiple turns, sometimes using self-correcting or planning strategies (e.g., Self-Edit, Self-Planning, Self-Debugging, Self-Collaboration). These approaches increase accuracy by allowing the model to revisit and adjust its output iteratively. 

\textbf{Multi-Turn Methods} include Self-Collaboration~\cite{dong2024selfcollaborationcodegenerationchatgpt}, AgentCoder\cite{huang2024agentcoder}, CodeCoT~
\cite{huang2024codecottacklingcodesyntax}, MapCoder~\cite{islam2024mapcodermultiagentcodegeneration} and LDB~\cite{zhong2024debuglikehumanlarge} provide structured interactions between multiple agents to collaboratively refine code solutions. These methods are designed to improve accuracy through the collective input of several agents rather than relying on a single response.

For our experiments, we select three representative and high-performing methods in the multi-agent approach: Self-Collaboration, AgentCoder, and LDB.

% By categorizing these methods, we can assess their performance across multiple coding benchmarks (HumanEval, HumanEval-ET) and evaluate the overall mean accuracy, demonstrating the effectiveness of different approaches in handling complex code generation and evaluation tasks.

\subsection{LLMs}
For LLMs, we use several widely adopted large language models, including GPT-3.5-Turbo, GPT-4o, o1-mini, Claude-3-Haiku, Claude-3.5-Sonnet, and DeepSeek-V3, ensuring the method's transferability across different models.

\textbf{ChatGPT.} Based on OpenAI's GPT architecture, GPT-3.5-Turbo offers faster processing and lower costs compared to GPT-3.5. GPT-4o-Mini is a smaller, cost-efficient variant of GPT-4, while GPT-4o provides strong language capabilities with lower computational demands. Additionally, the o1-mini model is a compact and lightweight version designed for scenarios requiring efficient resource utilization, delivering balanced performance with reduced computational overhead~\cite{radford2018improving, radford2019language, brown2020language}.

\textbf{Claude.} Developed by Anthropic, Claude focuses on safe and controllable AI interactions. Claude-3-Haiku is a specialized version designed for faster, efficient text generation and comprehension, making it ideal for real-time applications. Additionally, Claude-3.5-Sonnet is an advanced iteration that enhances the balance between performance and efficiency, offering improved contextual understanding and nuanced text generation, suitable for more complex and detailed tasks~\cite{bai2022constitutionalaiharmlessnessai, bai2022traininghelpfulharmlessassistant}.

\textbf{DeepSeek.} The DeepSeek series by DeepSeek Technologies excels in natural language processing and code-related tasks. The latest version, DeepSeek-V3, represents a significant advancement, integrating and enhancing the capabilities of its predecessors into a unified, state-of-the-art model. DeepSeek-V3 delivers superior general and coding performance, achieves better alignment with human preferences, and introduces optimizations for writing tasks and instruction-following scenarios. This version is designed to meet the growing demands of complex applications, offering improved efficiency, accuracy, and versatility~\cite{deepseekai2024deepseekv3technicalreport}.

% We use official APIs for all models and verify their versions and training data cut-off dates before and after the experiments to ensure consistency.

\subsection{Parameter Settings}
%% temperature in code generation and code translation
In the experiments, the temperature parameter is set to $0$ for both code generation and code translation, a common practice when employing large language models for such tasks~\cite{huang2024agentcoder, zhong2024debuglikehumanlarge, 10.1145/3697010}.
For the prompt mutation process, the temperature is set to $1.0$ to increase diversity and creativity, allowing for novel and varied prompt mutations.

%% trainset来源
The training set consists of $20$ data samples, with $10$ samples sourced from each of the two primary categories: (1) existing data, and (2) LLM-mutated data. For existing data, to ensure that they are distinct from the test set, we randomly sample $10$ samples from an external dataset. Specifically, for Code Generation tasks: HumanEval uses existing data from MBPP, MBPP uses data from HumanEval, and LiveCodeBench sources its existing data from AVATAR (code generation version). For Code Translation tasks: AVATAR uses CodeNet data, while CodeNet uses AVATAR as its existing data source. This cross-dataset curation ensures the independence of the test set while preserving task relevance.
For the mutation process, in each iteration, $10$ different mutated prompts are generated. This number is selected to introduce sufficient diversity into the prompt pool, enabling the exploration of various prompt structures and linguistic variations without incurring excessive computational costs. Generating $10$ mutated prompts per iteration strikes a balance between diversity and efficiency.

% \textbf{Iteration Termination Criteria. } The iterative prompt optimization process continues until one of two termination conditions is met: (1) a maximum of 10 iterations is reached, or (2) the maximum quantifiable score among the generated prompt sets remains the same for two consecutive iterations. The first condition ensures that the process does not continue indefinitely, preventing excessive computational resource usage. The second condition serves as a convergence criterion, indicating that the optimization process has plateaued and further iterations are unlikely to yield significant improvements.

For the parameters in the LLM API calls, including max tokens, top p, and others, are kept at their default settings as specified by each model.
When extracting code snippets from the model's results, we aim to prevent situations where multiple code segments in a single response could result in the extraction of non-executable code. To address this, we adopt the method proposed in paper~\cite{10.5555/3666122.3667065}, extracting the longest compilable code snippet from each model response.

\subsection{Evaluation Metric}
Following previous code generation studies~\cite{chen2022codet, nijkamp2022codegen, 10.1145/3690635}, we adopt Pass@1 as our evaluation metric for both generation and translation tasks. In this metric, the code generation model generates only one program in response to a given requirement. The requirement is considered solved if the generated program passes all predefined test cases. The Pass@1 score is then calculated as the percentage of requirements for which the model successfully generates a program that passes all the test cases on the first try. This score provides a direct measure of a model's ability to produce fully functional and correct code in a single attempt, reflecting both its precision and reliability in practical applications.
% The Pass@k metric measures the percentage of solved requirements out of the total, with higher values indicating better performance. In our experiments, we specifically set k = 1, specifically focusing on the Pass@1 metric to assess the model's effectiveness.

% The significance of Pass@1 lies in its ability to directly measure the model’s capacity to generate correct code in a single turn. A high Pass@1 value reflects the model's proficiency in producing accurate programs on the first try, which is important in practical applications where minimizing the number of attempts is crucial.

%% 无偏pass@k的描述
% Moreover, prior studies~\cite{athiwaratkun2022multi, chen2021evaluating} have demonstrated that the standard Pass@k metric can exhibit high variance, prompting the development of an unbiased Pass@k to enhance evaluation accuracy. Following these studies, we employ this unbiased Pass@k approach by generating n $\geq$ k programs per requirement, counting the number of solved requirements c, and subsequently calculating the unbiased Pass@k value.

% \begin{equation}
% \text{Pass@k} = \mathbb{E}_{\text{Problems}} \left[ 1 - \frac{\binom{n-c}{k}}{\binom{n}{k}} \right]
% \end{equation}

% It is also noteworthy that earlier code generation research often utilized text-similarity-based metrics (such as BLEU~\cite{papineni2002bleu} and CodeBLEU~\cite{ren2020codebleu}), however, these metrics have been shown to be inadequate in assessing the correctness of generated code~\cite{chen2021evaluating}. As such, we exclude these metrics from our experiments.

\subsection{Experimental Platform}

We conduct the experiments on an NF5468M6 server with two Intel Xeon Gold 6354 C processors and 512 GB of memory (16 x 32 GB DIMM DDR4). The server includes a 480 GB and a 24 TB logical volume for ample storage. It is equipped with eight NVIDIA RTX 4090 GPUs for high-performance computing and has eight Gigabit Ethernet interfaces for reliable data transmission.

\section{Results and Analysis}
\label{sec:results}
\subsection{RQ1: How does \Name{} perform compared to other methods?}
To address the first research question, we assess the performance of \Name{} on two types of tasks: natural-language-based code generation and code translation. 
%The code generation task entails producing new code from natural language descriptions, while the code translation task involves converting code from one programming language to another. Together, these tasks capture different facets of the model's code generation capabilities.

\begin{table*}[t]
\centering
\caption{Prochemy on Different Models (HumanEval and MBPP)}
\label{model results}
% \vspace{-2mm}
\resizebox{1.0\linewidth}{!}{ % 缩放表格到页面宽度
\begin{tabular}{lcccccccccccccccc}
\toprule
\multirow{2}{*}{\textbf{Prompting Methods}} & 
\multicolumn{4}{c}{\textbf{HumanEval}} & 
\multicolumn{4}{c}{\textbf{HumanEval+}} & 
\multicolumn{4}{c}{\textbf{MBPP}} & 
\multicolumn{4}{c}{\textbf{MBPP+}} \\
\cmidrule(lr){2-5} \cmidrule(lr){6-9} \cmidrule(lr){10-13} \cmidrule(lr){14-17}
 & \textbf{GPT-3.5} & \textbf{GPT-4o} & \textbf{Claude-3} & \textbf{DeepSeek-V3} & \textbf{GPT-3.5} & \textbf{GPT-4o} & \textbf{Claude-3} & \textbf{DeepSeek-V3} & \textbf{GPT-3.5} & \textbf{GPT-4o} & \textbf{Claude-3} & \textbf{DeepSeek-V3} & \textbf{GPT-3.5} & \textbf{GPT-4o} & \textbf{Claude-3} & \textbf{DeepSeek-V3} \\
\midrule

\textbf{Single-Turn} \\
Zero-Shot         & 72.6 & 90.2 & 76.8 & 90.2 & 68.9 & 81.7 & 68.9 & 84.8 & 75.9 & 86.5 & 80.2 & 87.6 & 65.3 & 72.5 & 68.8 & 73.0 \\
Few-shot          & 75.0 & 90.2 & 78.0 & 90.9 & 70.1 & 84.8 & 68.9 & 86.6 & 76.5 & 88.6 & 80.2 & 88.1 & 68.0 & 73.0 & 69.6 & 73.3 \\
CoT               & 75.6 & 91.5 & 76.2 & 89.0 & 70.1 & 85.4 & 67.1 & 82.9 & 78.8 & 87.8 & 79.4 & 87.8 & 65.6 & 73.4 & 68.5 & 71.7 \\
APE               & 73.2 & 90.2 & 77.4 & 90.2 & 68.9 & 83.5 & 69.5 & 85.3 & 78.6 & 88.6 & 80.7 & 87.3 & 64.0 & 73.0 & 69.0 & 72.5 \\
OPRO              & 75.0 & 89.0 & 74.4 & 89.6 & 69.5 & 83.5 & 67.7 & 86.0 & 78.8 & 88.9 & 80.4 & 86.8 & 65.9 & 73.3 & 67.7 & 72.8 \\
BPO               & 75.6 & 90.2 & 76.8 & 91.5 & 69.5 & 86.0 & 68.3 & 86.6 & 78.6 & 88.6 & 79.9 & 88.4 & 66.1 & 72.8 & 69.3 & 73.5 \\
Self-Collaboration-noiter & 74.4 & 90.2 & 77.4 & 90.9 & 68.9 & 86.0 & 69.5 & 86.6 & 80.2 & 88.1 & 80.7 & 91.5 & 67.5 & 72.8 & 69.6 & 77.5 \\
AgentCoder-noiter & 75.6 & 87.8 & 78.7 & 91.5 & 69.5 & 84.1 & 70.7 & 85.4 & 79.4 & 87.8 & 81.2 & 90.2 & 67.5 & 73.4 & 70.1 & 74.9 \\
LDB-noiter        & 76.2 & 91.5 & 78.0 & 91.5 & 70.7 & 84.8 & 70.1 & 85.4 & 80.7 & 89.7 & 81.0 & 90.7 & 68.3 & 75.4 & 69.3 & 75.1 \\
\textbf{Zero-shot + \Name{}} & \textbf{76.2} & \textbf{92.1} & \textbf{79.3} & \textbf{92.7} & \textbf{71.3} & \textbf{86.6} & \textbf{72.0} & \textbf{87.2} & \textbf{81.2} & \textbf{89.9} & \textbf{81.7} & \textbf{91.8} & \textbf{68.5} & \textbf{75.4} & \textbf{70.1} & \textbf{78.0} \\
\textbf{CoT + \Name{}}     & \textbf{77.4} & \textbf{92.7} & \textbf{80.5} & \textbf{93.3} & \textbf{72.0} & \textbf{89.3} & \textbf{72.6} & \textbf{88.4} & \textbf{82.5} & \textbf{89.9} & \textbf{82.0} & \textbf{92.3} & \textbf{69.0} & \textbf{75.7} & \textbf{70.9} & \textbf{78.3} \\

\midrule

\textbf{Multi-Turn} \\
Self-Collaboration& 78.0 & 90.9 & 81.1 & 92.1 & 70.1 & 87.8 & 71.3 & 87.2 & 82.5 & 88.6 & 82.3 & 91.8 & 69.0 & 74.1 & 70.1 & 77.5 \\
AgentCoder        & 79.3 & 92.7 & 80.5 & 93.3 & 73.2 & 87.2 & 72.0 & 85.4 & 82.2 & 88.9 & 82.8 & 91.5 & 69.3 & 75.1 & 71.1 & 76.7 \\
LDB               & 82.3 & 94.5 & 81.1 & 92.7 & 78.0 & 88.4 & 73.2 & 88.4 & 83.1 & 89.9 & 83.6 & 92.3 & 69.3 & 74.9 & 70.6 & 75.7 \\
\textbf{Self-Collaboration + Prochemy}   & \textbf{79.3} & \textbf{92.7} & \textbf{81.7} & \textbf{93.9} & \textbf{72.0}& \textbf{89.6} & \textbf{72.6} & \textbf{90.9} & \textbf{83.3} & \textbf{90.2} & \textbf{83.1} & \textbf{92.3} & \textbf{69.6} & \textbf{76.2} & \textbf{70.9} & \textbf{79.4} \\
\textbf{AgentCoder + Prochemy}    & \textbf{81.1} & \textbf{93.9} & \textbf{83.5} & \textbf{93.3} & \textbf{73.8} & \textbf{89.3} & \textbf{73.8} & \textbf{90.2} & \textbf{82.8} & \textbf{90.7} & \textbf{83.3} & \textbf{92.1} & \textbf{70.1} & \textbf{77.2} & \textbf{71.7} & \textbf{78.3} \\
\textbf{LDB + Prochemy}           & \textbf{83.5} & \textbf{96.3} & \textbf{84.8} & \textbf{94.5} & \textbf{79.3} & \textbf{90.2} & \textbf{73.8} & \textbf{90.9} & \textbf{83.9} & \textbf{92.6} & \textbf{84.1} & \textbf{93.1} & \textbf{70.4} & \textbf{80.7} & \textbf{71.4} & \textbf{78.8} \\

\bottomrule
\end{tabular}
}
\end{table*}

% 主模型结果表格
\begin{table*}[t]
\centering
\caption{Prochemy Performance on Different Models (LiveCodeBench)}
\resizebox{0.65\linewidth}{!}{
\begin{tabular}{lcccc}
\toprule
\textbf{Prompting Methods} & 
\multicolumn{4}{c}{\textbf{Models}} \\
\cmidrule(lr){2-5}
& \textbf{GPT-3.5-Turbo} & \textbf{GPT-4o} & \textbf{Claude-3.5-Sonnet} & \textbf{DeepSeek-V3} \\
\midrule
Zero-Shot & 10.9 & 22.8 & 12.9 & 24.8 \\
Few-shot & 6.9 & 21.8 & 11.9 & 24.8 \\
CoT & 11.9 & 20.8 & \textbf{16.8} & 22.8 \\
APE & 9.9 & 22.8 & 11.9 & 25.7 \\
OPRO & 9.9 & 20.8 & 13.9 & 23.8 \\
BPO & 10.9 & 22.8 & 11.9 & 20.8 \\
Self-Collaboration-noiter & 7.9 & 19.8 & 14.9 & 19.8 \\
AgentCoder-noiter & 11.9 & 19.8 & 7.9 & 16.8 \\
LDB-noiter & 7.9 & 17.8 & 7.9 & 14.9 \\
Zero-shot + \Name{} & \textbf{12.9} & \textbf{23.8} & \textbf{16.8} & \textbf{25.7} \\
CoT + \Name{} & \textbf{12.9} & \textbf{24.8} & \textbf{16.8} & \textbf{27.7} \\
\bottomrule
\end{tabular}
}
\label{different models results}
\end{table*}

\begin{table*}[t]
\centering
\caption{Prochemy on Different Models (Code Translation)}
% \vspace{-2mm}

\resizebox{1.0\linewidth}{!}{ % 缩放表格到页面宽度
\begin{tabular}{lcccccccccccccccc}
\toprule
\textbf{Prompting Methods} & 
\multicolumn{8}{c}{\textbf{Java2Python}} & 
\multicolumn{8}{c}{\textbf{Python2Java}} \\
\cmidrule(lr){2-9} \cmidrule(lr){10-17}
& \multicolumn{2}{c}{\textbf{GPT-3.5}} & \multicolumn{2}{c}{\textbf{GPT-4o}} & \multicolumn{2}{c}{\textbf{Claude-3-Haiku}} & \multicolumn{2}{c}{\textbf{DeepSeek-V3}} & 
\multicolumn{2}{c}{\textbf{GPT-3.5}} & \multicolumn{2}{c}{\textbf{GPT-4o}} & \multicolumn{2}{c}{\textbf{Claude-3-Haiku}} & \multicolumn{2}{c}{\textbf{DeepSeek-V3}} \\
\cmidrule(lr){2-3} \cmidrule(lr){4-5} \cmidrule(lr){6-7} \cmidrule(lr){8-9} 
\cmidrule(lr){10-11} \cmidrule(lr){12-13} \cmidrule(lr){14-15} \cmidrule(lr){16-17}
& \textbf{CodeNet} & \textbf{AVATAR} & \textbf{CodeNet} & \textbf{AVATAR} & \textbf{CodeNet} & \textbf{AVATAR} & \textbf{CodeNet} & \textbf{AVATAR} & 
\textbf{CodeNet} & \textbf{AVATAR} & \textbf{CodeNet} & \textbf{AVATAR} & \textbf{CodeNet} & \textbf{AVATAR} & \textbf{CodeNet} & \textbf{AVATAR} \\
\midrule
Zero-Shot & 60.5 & 63.8 & 63.5 & 74.5 & 59.0 & 63.0 & 72.0 & 80.9 & 76.5 & 54.4 & 95.5 & 66.8 & 86.0 & 51.5 & 91.0 & 87.7 \\
Few-shot & 69.5 & 64.4 & 70.0 & 76.2 & 60.5 & 64.4 & 82.0 & 81.6 & 81.0 & 54.9 & 96.0 & 68.1 & 87.5 & 53.2 & 94.0 & 88.5 \\
CoT & 63.0 & 62.8 & 63.5 & 73.2 & 58.5 & 61.5 & 64.0 & 78.7 & 81.0 & 51.9 & 96.5 & 64.7 & 86.5 & 49.8 & 92.0 & 86.0 \\
APE & 61.0 & 64.9 & 72.0 & 77.0 & 60.0 & 63.6 & 82.0 & 83.4 & 79.0 & 55.3 & 96.5 & 66.0 & 87.0 & 56.2 & 94.0 & 87.2 \\
OPRO & 61.0 & 63.8 & 69.5 & 78.7 & 59.5 & 64.9 & 70.5 & 80.3 & 81.5 & 54.4 & 96.5 & 67.7 & 87.0 & 51.9 & 93.5 & 86.8 \\
BPO & 60.5 & 65.1 & 64.5 & 63.8 & 60.0 & 60.9 & 67.5 & 79.6 & 80.0 & 56.9 & 95.5 & 65.5 & 86.0 & 46.8 & 91.0 & 85.7 \\
\textbf{Zero-shot + \Name{}} & \textbf{73.0} & \textbf{66.0} & \textbf{79.5} & \textbf{84.1} & \textbf{64.0} & \textbf{68.5} & \textbf{86.5} & \textbf{88.9} & \textbf{83.0} & \textbf{57.7} & \textbf{97.0} & \textbf{78.2} & \textbf{88.5} & \textbf{61.9} & \textbf{95.5} & \textbf{91.9} \\
\bottomrule
\end{tabular}
}
\label{code translatio results}
\end{table*}

\subsubsection{Natural-Language-Based Code Generation}
We evaluate natural language-based code generation across five benchmarks: HumanEval/MBPP (base functionality), their augmented versions HumanEval+/MBPP+ (strict correctness and robustness), and LiveCodeBench (dynamically refreshed challenges for advanced code generation).

We first introduce the experimental results of HumanEval, HumanEval+, MBPP, and MBPP+, where we select all baseline methods for evaluation. Subsequently, we provide the results of LiveCodeBench. Due to the complexity of the problems and the high computational costs associated with multi-round iterative approaches, we limited the baselines to include only non-iterative versions of single-round and multi-round methods(projected to exceed hundreds of hours for full multi-round evaluation). This choice is driven by the need to balance evaluation comprehensiveness with resource constraints.

% We separate LiveCodeBench results for two reasons. First, LiveCodeBench features significantly more complex problems requiring sophisticated algorithmic reasoning and intricate edge case handling. Second, our preliminary analysis shows each multi-round iteration for a single LiveCodeBench problem consumes over 150 seconds and 20K tokens per iteration – with challenging problems potentially requiring dozens of iterations. Given these costs (projected to exceed hundreds of hours for full multi-round evaluation) and the dataset's multidimensional I/O complexity exacerbating execution overhead, we restrict LiveCodeBench experiments to single-turn generation under our computational budget constraints. This ensures methodological consistency while maintaining practical feasibility.

\textbf{Analysis of \Name{} Across Models on HumanEval, HumanEval+, MBPP, and MBPP+.} 
The results of these datasets are shown in Table~\ref{model results}. 
We first examine the performance of \Name{} when integrated with zero-shot prompting. Across all datasets and models, Zero-shot + \Name{} consistently boosts performance, delivering an average gain of $4.04\%$. Notably, for the more challenging HumanEval+ and MBPP+ variants, \Name{} lifts average performance by $4.21\%$ and $4.43\%$, respectively. 
Such results also outperform existing prompt optimization approaches by an average of $2.33\%$. This average improvement is conservatively calculated by comparing the proposed method with the best-performing approach among APE, OPRO, and BPO for each model.
% APE raises GPT-4o on HumanEval+ by only 1.8\%, while Zero-shot + \Name{} achieves 4.9\%). These improvements remain stable across various model types, setting \Name{} apart from OPRO and BPO, which exhibit inconsistent behavior. For instance, OPRO degrades Claude-3’s HumanEval performance by 2.4\%, whereas \Name{} delivers a +2.5\% improvement.

Next, we explore the impact of \Name{} when combined with explicit chain-of-thought reasoning. Here, Prochemy provides an average improvement of $4.55\%$ compared with standalone CoT method. The synergy between Prochemy and CoT is especially evident in complex reasoning tasks such as HumanEval+ and MBPP+, which see gains of $5.52\%$ and $5.26\%$, respectively. This partnership also mitigates the occasional instability of standalone CoT methods, which can sometimes decrease performance (e.g., Claude-3’s baseline CoT). By adaptively refining the logic in prompts, Prochemy ensures that the reasoning steps remain coherent. As a concrete example, GPT-4o’s HumanEval+ performance rises by $7.6\%$ when CoT is augmented by Prochemy—an improvement $2.45×$ greater than using CoT alone.

Finally, we investigate \Name{}’s capacity to optimize multi-turn code generation approaches such as Self-Collaboration, AgentCoder, and LDB. In these pipelines, LLMs typically draw on domain knowledge over several iterations, but the initial and intermediate prompts can still be refined. In our experiments, we employ \Name{} to optimize the initial prompt of them, we observe an average improvement of $2.00\%$. HumanEval-series tasks experience the greatest gains ($+2.23\%$), followed by MBPP-series benchmarks ($+1.77\%$). This iterative optimization helps reduce error propagation, ultimately improving the seed program’s quality. Notably, \Name{}-enhanced LDB establishes new state-of-the-art results across all benchmarks, including $96.3\%$ (GPT-4o on HumanEval), $90.9\%$ (DeepSeek-V3 on HumanEval+), and $93.1\%$ (DeepSeek-V3 on MBPP). These results underscore \Name{}’s strength in refining dynamic prompts for complex multi-stage workflows.

Beyond the above observations, it is particularly noteworthy that \Name{} delivers strictly positive gains for all tested models and datasets without any performance regressions. The consistent upward trend further validates \Name{}’s robustness and reliability as a prompting optimization framework.

\textbf{Analysis of \Name{} Across Models on LiveCodeBench.} 
We further validate \Name{}’s transferability on the more challenging LiveCodeBench dataset, as presented in Table~\ref{different models results}. Notably, only single-turn code generation methods are explored here, in part due to cost constraints and also to focus on how \Name{} performs under more direct prompting scenarios. Additionally, the use of a higher-capacity Claude-3.5-Sonnet model reflects the increased difficulty of LiveCodeBench, enabling a clearer assessment of \Name{}’s robustness across varying model strengths.

For all models tested, \Name{} strengthens the baseline Zero-Shot results, with especially large gains for less advanced architectures. After applying the \Name{} method, the average performance improved by $14.15\%$ compared to the zero-shot method. Claude-3.5-Sonnet, for example, improves substantially from $12.9$ to $16.8$ ($+30.2\%$), emphasizing \Name{}’s ability to refine underperforming baselines. Meanwhile, DeepSeek-V3 and GPT-4o shows a smaller but still positive improvement ($24.8$ \(\rightarrow\) $25.7$, $+3.6\%$, and $22.8$ \(\rightarrow\) $23.8$, $+4.3\%$).

Beyond Zero-Shot settings, \Name{} also enhances reasoning-oriented methods, achieving an average improvement of $12.28\%$. For GPT-4o, combining Prochemy with chain-of-thought reasoning yields a pass@1 of $24.8$, translating to a $19.2\%$ gain over standalone CoT ($20.8$). DeepSeek-V3 demonstrates a similar pattern, where CoT + \Name{} achieves $27.7$, representing a $21.5\%$ jump over vanilla CoT ($22.8$). These results affirm that \Name{} can often amplify structured reasoning, especially in models with well-aligned CoT capabilities.

Across all tested architectures, \Name{} consistently surpasses specialized prompt optimization approaches like APE, OPRO, and BPO. For example, on GPT-4o, Zero-Shot + \Name{} ($23.8$) outperforms APE ($22.8$), OPRO ($20.8$), and BPO ($22.8$), underscoring the effectiveness of systematic prompt refinement over hand-tuned strategies. These findings illustrate that \Name{}’s improvements hold even on advanced datasets such as LiveCodeBench.

\subsubsection{Code Translation}
We further extend the evaluation of \Name{} to code translation tasks, systematically investigating its efficacy in code-to-code generation scenarios beyond conventional natural language-to-programming language settings. Table~\ref{code translatio results} demonstrates \Name{}’s performance across models and datasets.

Similar results are observed in code generation tasks, \Name{} boosts Zero-Shot performance, achieving an average improvement of $13.68\%$ in Java-to-Python translation tasks, And in Python-to-Java task, this value is $8.25\%$. For instance, in Java-to-Python translation on the AVATAR dataset, GPT-4o scores $84.1$ under Zero-Shot + \Name{}, representing a $9.6\%$ absolute gain over standalone Zero-Shot ($74.5$). These results underscore how \Name{}’s adaptive refinement strategy enhances clarity and precision, even in cross-lingual code settings.

Across multiple datasets and models, \Name{} outperforms specialized prompt optimization techniques such as APE, OPRO, and BPO. In DeepSeek-V3’s Java-to-Python task on CodeNet, \Name{} + Zero-Shot achieves 86.5, notably exceeding APE’s $82.0$ and OPRO’s $70.5$. Moreover, BPO lags further behind at $67.5$. This performance gap aligns with findings in other code generation benchmarks (e.g., HumanEval+ and MBPP+), where \Name{} consistently balances specificity and flexibility without over-constraining the output.

The results in code translation extend the evidence of \Name{}’s versatility across diverse programming tasks. By dynamically adjusting prompt structures to accommodate distinct source–target requirements, \Name{} effectively generalizes beyond single-language code generation and natural-language-to-code mapping. This adaptability not only cements \Name{} as a robust solution for a wide range of code generation challenges but also highlights its potential as a unifying prompting framework, bridging the gap between specialized transformations and more generic LLM-driven development workflows.

\finding{\textbf{Answer to RQ1:} \Name{} delivers consistent performance gains across code generation and translation tasks. For code generation, it achieves $+4.04\%$ (zero-shot), $+4.55\%$ (CoT), and $+2.00\%$ (multi-turn) average improvements over baselines, with HumanEval+/MBPP+ seeing $+4.21–7.76\%$ gains. It sets new state-of-the-art results (e.g., $96.3\%$ on HumanEval) through LDB integration. On LiveCodeBench, \Name{} boosts zero-shot and CoT performance by $14.15\%$ and $12.28\%$. In code translation, it achieves $+13.68\%$ average improvement on Java-Python tasks, and $+8.25\%$ on Python-Java tasks. These results validate \Name{} as a unified, efficient framework for code-related LLM optimization.}

\subsection{RQ2: What effects do each component in the \Name{} have?}

\begin{table*}[t]
\centering
\caption{Ablation Study of Iteration Effects on \Name{}'s Performance}
% \vspace{-2mm}

\resizebox{0.5\linewidth}{!}{ % 缩放表格到页面宽度
\begin{tabular}{lcccc}  % 确保有五个列对齐字符
\toprule
\textbf{Models} & 
\textbf{\begin{tabular}[c]{@{}c@{}}Zero-Shot\end{tabular}} & \textbf{\begin{tabular}[c]{@{}c@{}}w/o Iter\end{tabular}} &
\textbf{\begin{tabular}[c]{@{}c@{}}Iter\_10\end{tabular}} & \textbf{\begin{tabular}[c]{@{}c@{}}Prochemy\end{tabular}} \\ \midrule
GPT-3.5-Turbo & 72.6 & 73.8 & 75.0 & \textbf{76.2} \\ 
GPT-4o & 86.0 & 89.6 & 90.2 & \textbf{92.1} \\ 
Claude-3-Haiku & 73.7 & 76.2 & 77.4 & \textbf{78.7} \\
DeepSeek-V3 & 87.2 & 87.8 & 89.6 & \textbf{90.2} \\ 

\bottomrule
\end{tabular}
}

\label{abalation1}
\end{table*}

\begin{table*}[t]
\centering
\caption{Ablation Study of Weighted Score Effects on \Name{}'s Performance}
\resizebox{0.8\linewidth}{!}{ % 缩放表格到页面宽度
\begin{tabular}{l c c c c c c}
\toprule
\textbf{Models} & \textbf{w/o WS} & \textbf{\Name{}} & \textbf{Reduction (\%)} & \textbf{w/o WS} & \textbf{\Name{}} & \textbf{Improvement (\%)}\\
\cmidrule{2-7}
\textbf{} & Iterations & Iterations & Iterations & Pass@1 & Pass@1 & Pass@1\\
\midrule
GPT-3.5-Turbo    & 5.2 & \textbf{4.0} & 23.1\% &  70.7 & \textbf{76.2} & 3.5\% \\
GPT-4o           & 4.2 & \textbf{3.3} & 21.4\% &  90.2 & \textbf{92.1} & 2.1\% \\
Claude-3-Haiku   & 5.4 & \textbf{4.3} & 20.4\% &  73.1 & \textbf{78.7} & 7.7\% \\
DeepSeek-V3      & 4.4 & \textbf{3.6} & 18.2\% &  87.2 & \textbf{90.2} & 3.4\% \\
\bottomrule
\end{tabular}
}
\label{abalation2}
\end{table*}

%% 消除哪些组件，为什么要消除这些组件
To thoroughly assess the contributions of individual components within \Name{}, we conduct ablation experiments by systematically removing specific elements of the iterative process. Specifically, we examine the effects of (1) eliminating the iterative optimization to understand its impact on model performance, (2) using a fixed number of iterations (e.g., 10 iterations) to evaluate the necessity of adaptive iteration counts with early stopping, and (3) removing the weighted scoring mechanism used for selecting prompts during iterations to determine its role in accelerating convergence and enhancing performance. By isolating and analyzing these components, we aim to understand their individual contributions to the overall effectiveness of \Name{} in code generation tasks. In particular, we conduct this ablation test on the HumanEval dataset due to the cost.

\subsubsection{Iteration}
To evaluate the impact of the iterative process in \Name{}, we conduct experiments under different iteration settings.

\paragraph{No Iteration}
We begin by conducting experiments without iterative processing, as shown in Table~\ref{abalation1}. A comparison between the non-iterative (w/o Iter) and original versions of \Name{} indicates that the removal of the iterative procedure generally reduces the method's effectiveness in improving model performance. This decline may be attributed to the non-iterative method's limited ability to explore different directions, constraining its capacity to identify the most effective prompts for the given tasks.

\paragraph{Fixed Iteration Count of 10}
We also conduct experiments with a fixed iteration count of $10$. This choice is informed by preliminary findings, which suggest that at this number of iterations, most models have already produced results that meet the early stopping criteria.

We evaluate the performance of each model on the HumanEval dataset using both the early stopping method (\Name{}) and the 10-iteration approach (Iter\_10). The results, as shown in Table~\ref{abalation1}, demonstrate that \Name{} using early stopping consistently outperforms the version with a fixed number of iterations. 
On average, models show an improvement of around $1.50\%$ in pass@1 scores when using early stopping compared to the 10-iteration approach. This trend is consistent across all models, suggesting that the early stopping method optimizes prompt refinement more effectively and efficiently than extending iterations without significant additional gains.

When comparing the fixed 10-iteration approach (Iter\_10) to the non-iterative version (w/o Iter), it is evident that additional iterations lead to notable improvements in model performance, with an average increase of 1.48\% across models. This trend confirms that iterative refinement effectively enhances the models’ ability to generate more accurate and contextually relevant code.

\subsubsection{Weighted Score}
Another ablation experiment aims to evaluate the effectiveness of the weighted scoring mechanism in \Name{}.

In the w/o WS (without weighted scoring) setting, the weighting mechanism is not applied; instead, the original pass@1 values are directly used to select the best prompt for each iteration. This approach primarily affects both the performance of \Name{} and the number of iterations required to optimize the prompts effectively. Without the weighted scoring, which prioritizes prompts based on their ability to handle complex tasks, the optimization process may take more iterations to converge to an optimal prompt, as each prompt is evaluated uniformly regardless of task complexity. This can lead to less efficient prompt refinement and potentially increase the computational effort required to achieve similar performance levels that are attained more swiftly with weighted scoring.

The results are shown in Table ~\ref{abalation2}. The first two columns represent the number of iterations required to reach the convergence condition without and with the introduction of the weighted averaging mechanism, whereas the last two columns represent the pass@1 scores.

We observe that \Name{} consistently outperforms the version without weighted scoring (w/o WS) in terms of pass@1 scores. The integration of \Name{}'s weighted scoring mechanism results in an average improvement of $4.2\%$ in pass@1 scores across models, showcasing substantial enhancement. This highlights the benefits of employing such a mechanism.

Additionally, across all models, fewer iterations are required to converge when using the weighted scoring mechanism compared to the unweighted setting. On average, models require $20.8\%$ fewer iterations with weighted scoring compared to the w/o WS setting. This reduction in iterations is consistent across different models, highlighting the efficiency of the weighted scoring mechanism.

These findings emphasize that the inclusion of weighted scoring facilitates a more targeted and efficient optimization process. It not only reduces the number of iterations needed but also enhances the final output quality, streamlining the prompt refinement process effectively.

\finding{\textbf{Answer to RQ2: }Ablation studies show that each component of \Name{} enhances its performance. Iterative processing improves pass@1 scores, as non-iterative methods underperform. The early stopping mechanism and the weighted scoring mechanism accelerate convergence while improving the final output. }

%% rq3 Can our method be combined with existing other methods to achieve better results

\subsection{RQ3: In what key aspects and to what extent does \Name{} improve models' code capabilities compared to previous methods?}

To evaluate the improvements of the proposed \Name{} over traditional prompting techniques in code generation within a more specific context, we conduct a case study based on the results of the GPT-3.5-Turbo model.

\begin{figure*}[t]
% \vspace{-0.2cm}
\centering
\includegraphics[width=0.95\textwidth]{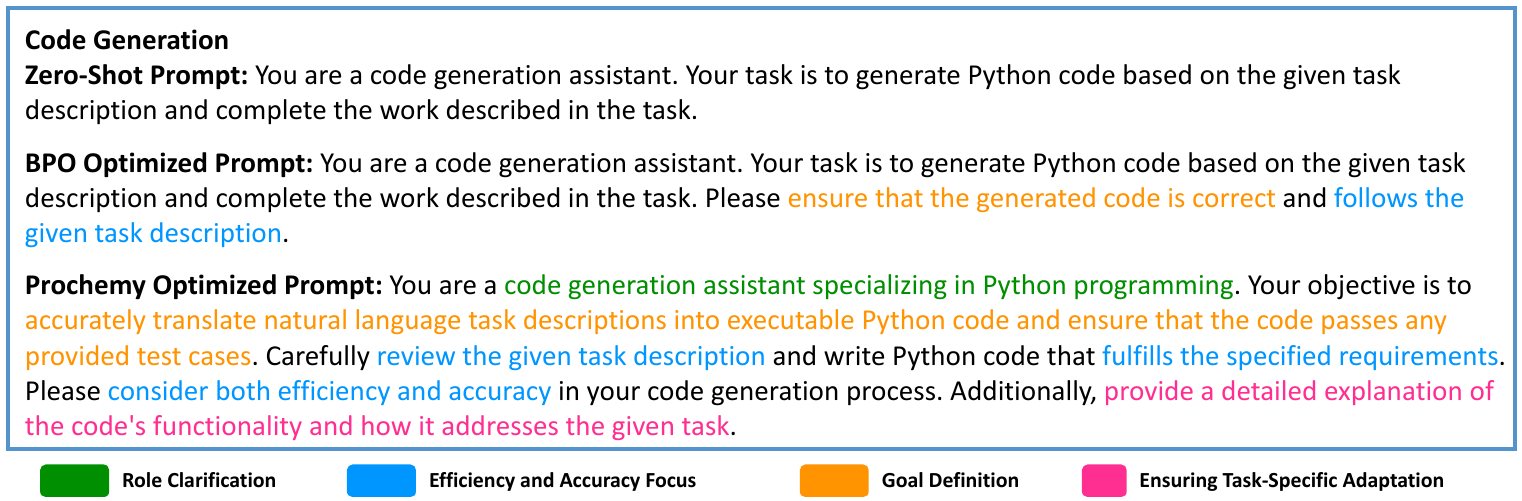}
% \vspace{-0.4cm}
\caption{Case 1 for Optimized Prompt}
% \vspace{-0.3cm}
\label{case3}
\end{figure*}

\subsubsection{Interpretability of \Name{}}
% Compared to alignment methods based on model training, such as Proximal Policy Optimization (PPO)~\cite{schulman2017proximalpolicyoptimizationalgorithms} or Direct Preference Optimization (DPO)~\cite{rafailov2024directpreferenceoptimizationlanguage}, a
A key advantage of \Name{} lies in its enhanced interpretability. By directly comparing the instructions before and after optimization, we can explicitly examine the mechanisms through which \Name{} operates. To analyze and generalize the optimization patterns of \Name{}, we review 300 samples generated during training and identified several recurring optimization strategies.

As illustrated in Figure \ref{case3}, using a concrete example of prompt optimization within a code generation task, we identify four frequently employed optimization strategies within \Name{}: Role Clarification, Goal Definition, Efficiency and Accuracy Focus, and Ensuring Task-Specific Adaptation. These strategies are not mutually exclusive in practical use, and this example represents a typical instance where all four are applied concurrently.

\begin{itemize}
    \item \textbf{Role Clarification} ensures that the model remains focused on the current software engineering task, rather than responding to general inquiries. This focus enables the model to fully utilize its coding capabilities acquired during pre-training.

    \item \textbf{Goal Definition} specifies the ultimate objective of the current task, guiding the assistant to prioritize accuracy and correctness in code generation, thereby ensuring that the resulting solution satisfies the task requirements.

    \item \textbf{Efficiency and Accuracy Focus} emphasizes not only correctness but also the optimization of code performance. By applying this strategy, the model ensures that the generated code maintains logical integrity, even in edge cases or under complex testing conditions.

    \item \textbf{Ensuring Task-Specific Adaptation} encourages the model to take into account the unique nuances and requirements of the specific task. During simulated gradient descent, feedback from the training set enables the model's prompts to gradually converge towards the optimal continuous vector space. In the context of discrete vector representations, this process manifests as task-specific adaptation. The specific implementation of this strategy varies considerably based on the requirements of the task.
\end{itemize}

In comparison, prompt optimization methods designed for natural language tasks, such as BPO~\cite{cheng2024blackboxpromptoptimizationaligning}, are limited to addressing only certain aspects of these strategies when applied to code-related tasks.

%% code generation
\subsubsection{Case for Code Generation}

\begin{figure*}[t]
\centering
% \vspace{-0.3cm}
\includegraphics[width=0.9\textwidth]{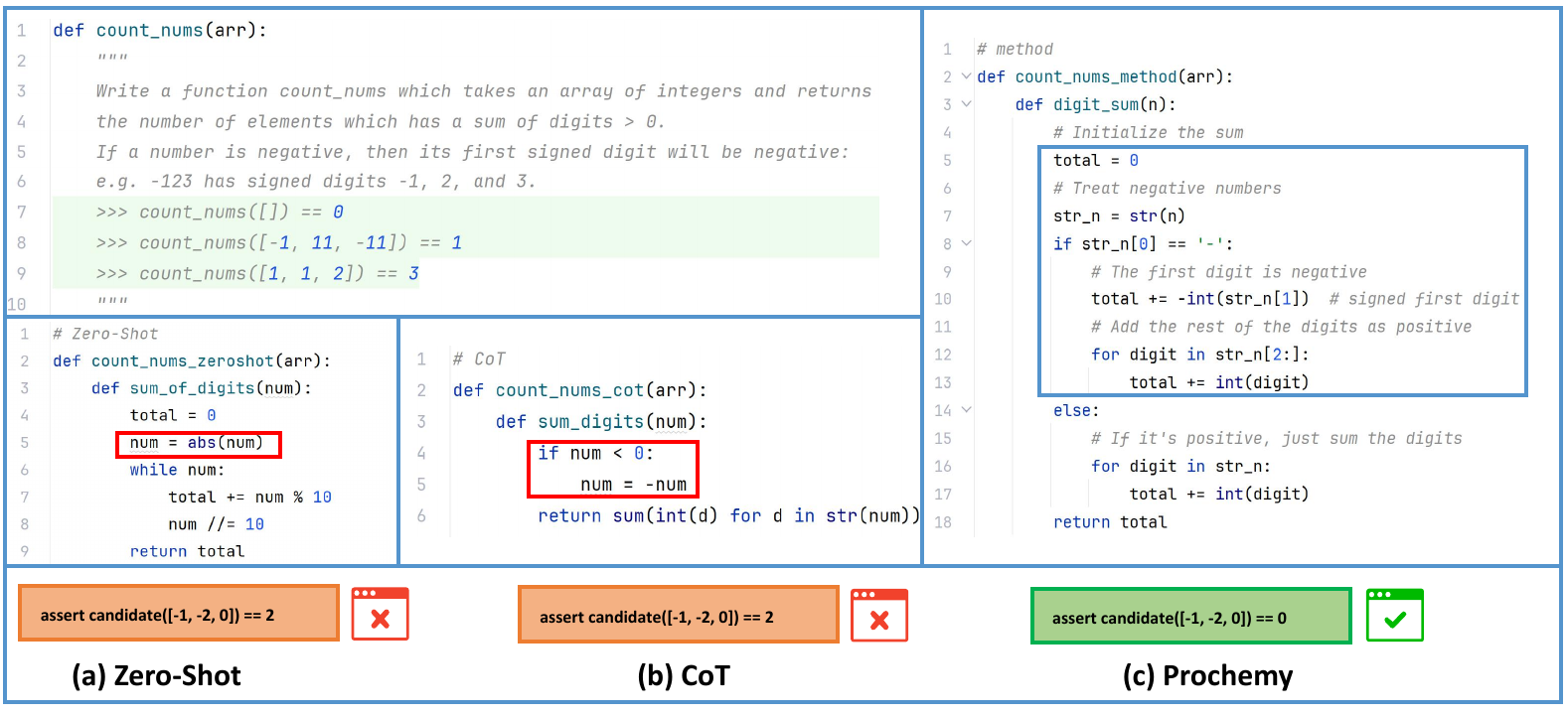}
% \vspace{-0.4cm}
\caption{Case 2 HumanEval/108
 for Code Generation}
% \vspace{-0.4cm}
\label{case1}
\end{figure*}

\begin{figure*}[t]
\centering
\includegraphics[width=0.85\textwidth]{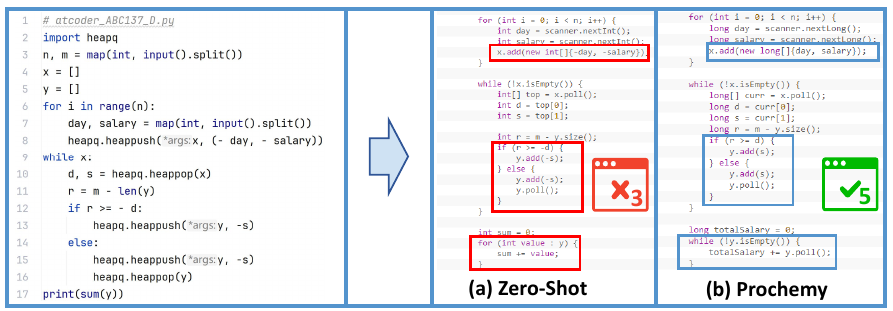}
% \vspace{-0.35cm}
\caption{Case 3 atcoder\_ABC137\_D for Code Translation}
% \vspace{-0.5cm}
\label{case2}
\end{figure*}

Fig.~\ref{case1} shows the code generation results for HumanEval/108 using different prompting methods. The Zero-Shot and Chain of Thought (CoT) methods generate incorrect results, while \Name{} produces the correct answer.

The task requires implementing a function, count\_nums, which takes an array of integers and returns the number of elements where the sum of their digits is greater than zero. A key detail is that for negative integers, the leading digit should be treated as negative.

The problem analysis clarifies why the zero-shot and CoT methods yield incorrect results. In the count\_nums\_zeroshot function, the sum\_digits function incorrectly treats negative numbers by converting them to their absolute values (num = abs(num)) before summing the digits. Thus, for input [-1, -2, 0], it computes -1 and -2 as 1 and 2, leading to errors.

Similarly, the CoT's result mishandles negative numbers by taking the absolute value (num = -num) and summing the digits as positive. This causes the same issue, incorrectly calculating -1 and -2 as 1 and 2 for the input [-1, -2, 0].

In contrast, the code generated by \Name{} produces correct results by using string manipulation to check if a number is negative (str\_n[0] == '-'). It treats the leading digit as negative and sums the remaining digits as positive, adhering to the problem's requirements. This approach shows that \Name{} better understands the problem specifications. Additionally, the code uses conditional logic, allowing it to handle inputs of varying lengths and types, accurately computing the sum of digits and their signs for both positive and negative integers.

This approach remains robust for integers of any magnitude, correctly handling all valid inputs—even with long integers and complex edge cases from the enhanced HumanEval-ET dataset. Additionally, code generated by \Name{} often includes crucial comments that enhance readability by explaining the logic and intent of each step, making the code's structure clearer and more intuitive—features less common in zero-shot and CoT results.

%% code translation
\subsubsection{Case for Code Translation}
In Figure \ref{case2}, we conduct an analysis of the code translation task using atcoder\_ABC137\_D as a case study. The task involves implementing the function max\_total\_reward, which takes three inputs: the number of tasks (N), the number of days (M), and two datasets (Ai and Bi). Ai represents the days after which the reward Bi can be received if the task is completed. The goal is to return the maximum total reward obtainable within M days.

The original Python code uses the ``heapq'' module to manage tasks in a min-heap (x), ordered by the latest completion day and reward. To simulate max-heap behavior, it stores negative reward values. A second min-heap (y) holds rewards for eligible tasks, with a size limit of M days.

The algorithm checks each task's latest completion day (d). If it can be completed within the remaining days (r), its reward is added to heap y. If not, the reward is added, and the smallest is removed to keep only the highest rewards. This ensures that the maximum total reward is selected at each step.

The Zero-Shot method produces incorrect results due to critical issues in code translation. The original Python code simulates a max-heap using ``heapq'' with negative values, but this is mishandled in the Java version. In Java, PriorityQueue is a min-heap by default, requiring either an explicit comparator or negative values to simulate a max-heap, both of which are missing in the Java code.

The Java code's custom comparator for \texttt{PriorityQueue<int[]>} mishandles negative values, unlike the original Python code that uses \texttt{-day} and \texttt{-salary} to simulate a max-heap properly. While the comparator in Java considers negative values, it does not apply them consistently for task selection and reward comparison. This leads to a logical inconsistency with the condition \( r \geq -d \); unlike in Python, the Java version fails to ensure tasks can be completed within the remaining days, resulting in incorrect task eligibility determination.

The Java implementation by zero-shot fails to replicate the Python code's core max-heap functionality for selecting optimal tasks based on rewards, leading to an inconsistencies and incorrect results. In contrast, \Name{}'s code is more reliable and logically consistent. It utilizes two PriorityQueues: one to store tasks sorted by descending completion day and reward, and another as a min-heap for the rewards of eligible tasks. An improvement is the use of the long type for day and salary variables, preventing overflow in large number operations and ensuring the code can handle larger datasets safely and efficiently.

\finding{
\textbf{Answer to RQ3: }\Name{} enhances model code capabilities by improving interpretability through strategies like Role Clarification and Goal Definition, optimizing prompts, and simplifying debugging. Case studies show \Name{} generates accurate, well-structured code, advancing the programming capabilities of large language models.
}

\section{Discussion}
\label{sec:discussion}

\begin{table}[t]
\centering
\caption{Prochemy Performance on o1-mini Model (LiveCodeBench)}
\resizebox{0.8\linewidth}{!}{ % 调整为45%页面宽度
\begin{tabular}{@{}p{0.6\linewidth}c@{}}
\toprule
\textbf{Prompting Methods} & \textbf{o1-mini} \\
\midrule
Zero-Shot          & 40.6 \\
Few-shot           & 38.6 \\
CoT                & 36.6 \\
APE                & 43.6 \\
OPRO               & 41.6 \\
BPO                & 41.6 \\
Self-Collaboration-noiter & 37.6 \\
AgentCoder-noiter  & 34.7 \\
LDB-noiter         & 41.6 \\
Zero-shot + \Name{} & \textbf{44.6} \\
CoT + \Name{}      & \textbf{43.6} \\
\bottomrule
\end{tabular}
}
\label{o1mini_results}
\end{table}

%HumanEval and MBPP 
\begin{table*}[t]
\centering
\caption{Time and Cost Analysis (HumanEval \& MBPP) }
\resizebox{1.0\linewidth}{!}{
\begin{tabular}{lcccccccccc}
\toprule
\textbf{Prompting Methods} & 
\multicolumn{4}{c}{\textbf{HumanEval}} & 
\multicolumn{4}{c}{\textbf{MBPP}} \\
\cmidrule(lr){2-5} \cmidrule(lr){6-9}
& \textbf{GPT-3.5-Turbo} & \textbf{GPT-4o} & \textbf{Claude-3-Haiku} & \textbf{DeepSeek-V3} & 
\textbf{GPT-3.5-Turbo} & \textbf{GPT-4o} & \textbf{Claude-3-Haiku} & \textbf{DeepSeek-V3} \\
\midrule
\textbf{Single-Turn} \\
Zero-Shot & 184+358/3.76s & 184+526/8.68s & 188+372/3.82s & 185+544/8.39s & 124+191/2.14s & 124+354/4.13s & 126+273/3.14s & 124+374/4.18s \\
Few-shot & 303+425/2.59s & 314+484/5.69s & 322+437/2.61s & 327+487/5.36s & 217+254/2.16s & 242+316/2.30s & 237+285/3.23s & 251+364/2.73s \\
CoT & 204+423/3.91s & 203+632/11.39s & 209+447/4.34s & 208+655/9.76s & 143+246/2.85s & 144+401/4.15s & 144+323/3.51s & 143+442/5.13s \\
APE & 281+482/3.46s & 314+705/5.53s & 339+497/4.60s & 322+712/9.78s & 197+333/2.25s & 242+495/4.28s & 268+414/3.77s & 295+515/4.78s \\
OPRO & 244+402/3.52s & 263+657/5.58s & 310+513/4.87s & 339+654/7.67s & 241+248/2.03s & 258+464/5.51s & 311+356/3.27s & 314+482/5.32s \\
BPO & 223+340/2.86s & 277+549/8.09s & 323+504/4.42s & 276+683/8.04s & 186+229/2.50s & 237+335/3.21s & 244+282/2.86s & 241+347/3.77s \\
Self-Collab-noiter & 235+403/2.04s & 236+551/5.26s & 241+411/2.92s & 238+566/6.30s & 177+160/1.56s & 178+278/1.93s & 180+219/1.75s & 182+281/1.87s \\
AgentCoder-noiter & 304+576/4.93s & 307+749/14.38s & 309+586/4.95s & 308+776/10.43s & 248+404/3.13s & 250+582/6.17s & 248+493/4.65s & 254+619/5.86s \\
LDB-noiter & 322+486/3.79s & 329+870/6.60s & 324+533/4.92s & 326+884/11.32s & 259+303/2.56s & 258+317/2.93s & 260+310/2.78s & 258+420/3.37s \\
Zero-shot+\Name{} & 259+483/2.79s & 312+641/11.45s & 317+503/4.76s & 322+702/10.08s & 251+339/2.94s & 307+512/4.23s & 295+463/3.28s & 337+462/3.91s \\
CoT+\Name{} & 294+462/3.51s & 355+728/11.73s & 342+542/5.01s & 357+743/10.94s & 287+392/3.02s & 339+537/4.40s & 311+479/4.07s & 362+499/4.38s \\

\midrule
\textbf{Multi-Turn} \\
Self-Collaboration & 407+813/10.42s & 492+1187/16.06s & 434+891/12.37s & 474+1171/19.28s & 447+541/14.17s & 527+672/18.36s & 428+577/15.62s & 547+880/24.01s \\
AgentCoder & 386+962/11.77s & 453+1011/14.44s & 408+994/11.96s & 458+1059/17.41s & 477+492/12.64s & 486+511/16.07s & 473+519/13.37s & 496+821/20.11s \\
LDB & 587+1094/14.49s & 523+1487/20.91s & 548+1169/16.77s & 515+1495/30.87s & 527+814/18.31s & 589+1066/22.16s & 559+798/18.76s & 541+974/29.55s \\
Self-Collab+\Name{} & 388+844/10.36s & 483+1172/15.49s & 427+906/12.44s & 490+1202/19.64s & 436+550/14.43s & 586+641/18.82s & 419+568/15.27s & 532+857/22.17s \\
AgentCoder+\Name{} & 443+1011/11.91s & 466+1064/14.61s & 489+1071/14.74s & 462+1097/18.02s & 490+516/13.02s & 527+549/16.88s & 478+551/13.84s & 507+864/20.98s \\
LDB+\Name{} & 467+1057/11.34s & 504+1390/20.03s & 511+1084/16.48s & 501+1447/29.64s & 499+786/17.18s & 591+1129/22.84s & 513+762/17.84s  & 588+942/29.38s \\

\bottomrule
\end{tabular}
}
\label{tab:humaneval_mbpp}
\end{table*}

% LiveCodeBench 结果表格
\begin{table*}[t]
\centering
\caption{Time and Token Cost Analysis (LiveCodeBench) }
\resizebox{0.8\linewidth}{!}{
\begin{tabular}{lccccc}
\toprule
\textbf{Prompting Methods} & \multicolumn{5}{c}{\textbf{LiveCodeBench}} \\
\cmidrule(lr){2-6}
& \textbf{GPT-3.5-Turbo} & \textbf{GPT-4o} & \textbf{Claude-3.5-Sonnet} & \textbf{DeepSeek-V3} & \textbf{o1-mini} \\
\midrule
\textbf{Single-Turn} \\
Zero-Shot & 544+370/4.28s & 533+663/19.01s & 584+833/13.31s & 525+764/11.47s & 587+5947/42.93s \\
Few-shot & 626+278/3.65s & 626+595/12.85s & 693+819/12.80s & 620+760/11.53s & 691+3982/25.96s \\
CoT & 540+388/4.39s & 540+706/19.79s & 591+846/13.99s & 532+752/11.37s & 594+5308/37.21s \\
APE & 609+388/4.53s & 609+714/15.42s & 666+842/15.56s & 600+830/16.25s & 666+4522/45.38s \\
OPRO & 546+396/4.86s & 546+702/15.96s & 599+852/14.39s & 538+812/13.35s & 594+5534/52.99s \\
BPO & 548+328/3.99s & 548+650/13.47s & 599+814/12.83s & 540+731/11.33s & 605+5551/51.95s \\
Self-Collab-noiter & 658+250/3.24s & 659+243/7.57s & 716+291/7.14s & 650+258/6.56s & 723+2473/15.60s \\
AgentCoder-noiter & 548+583/5.99s & 548+751/20.22s & 602+845/14.57s & 540+934/15.87s & 601+4981/44.20s \\
LDB-noiter & 548+456/4.87s & 517+646/15.24s & 568+845/13.36s & 509+964/17.55s & 560+8684/76.52s \\
Zero-shot+\Name{} & 589+532/5.83s & 587+745/19.58s & 641+837/13.86s & 580+830/12.79s & 645+5827/41.60s \\
CoT+\Name{} & 662+436/6.62s & 662+750/17.34s & 722+989/14.03s & 651+834/18.38s & 727+4561/41.13s \\

\bottomrule
\end{tabular}
}
\label{tab:livecodebench}
\end{table*}

\subsection{\Name{} on Reasoning Models (o1-mini)}
In this discussion, we include o1-mini in our experiments both because it is a high-performance model reportedly equipped with built-in chain-of-thought capabilities and because OpenAI’s recent guidelines suggest that elaborate prompt engineering may offer less benefit for such advanced reasoning models~\cite{openai2024o1}. However, the results in Table~\ref{o1mini_results} (pass@1 column for o1-mini) show that \Name{} still contributes measurable improvements even for o1-mini.

In single-turn settings, zero-shot prompting achieves a pass@1 of $40.6$, whereas few-shot ($38.6$) and CoT ($36.6$) yield lower scores. Adding \Name{} leads to zero-shot+\Name{} at $44.6$, which not only outperforms well-known techniques such as APE ($43.6$) and LDB-noiter ($41.6$) but also achieves the best overall pass@1 on o1-mini. This result indicates that \Name{}’s refinement mechanism can enhance code generation quality, even when the underlying model is designed for chain-of-thought reasoning. Interestingly, CoT+\Name{} ($43.6$) also boosts performance relative to standard CoT but is still slightly below zero-shot+\Name{}, suggesting that a concise prompting strategy—one that \Name{} inherently facilitates—can be more effective for o1-mini. This behavior aligns with OpenAI’s technical report, which notes that reasoning-oriented models like o1-mini often respond best to succinct prompts.

Overall, our findings demonstrate that o1-mini can still benefit from systematic prompt refinements via \Name{}, despite its built-in chain-of-thought capability and official recommendations favoring minimal prompt engineering. This underscores the versatility of \Name{}: it not only elevates performance in standard LLMs but also proves valuable for next-generation reasoning models.

% To investigate Prochemy's efficacy on modern models with inherent chain-of-thought (CoT) capabilities, we evaluate the o1-mini on LiveCodeBench, as results shown in Table~\ref{tab:livecodebench}. Our results reveal three key insights. First, while contemporary methods like APE (43.6) and LDB-noiter (41.6) show competitive performance, our proposed Prochemy method achieves superior results when combined with zero-shot prompting (44.6). Second, despite o1-mini's built-in chain-of-thought capabilities, applying Prochemy to CoT prompts still yields a 7-point improvement (43.6), demonstrating our method's complementary value to existing reasoning mechanisms. Third, we observe a distinctive pattern where zero-shot + Prochemy outperforms CoT + Prochemy (44.6 vs 43.6), contrasting with trends observed in other models. This aligns with OpenAI's technical report findings that reasoning-optimized models like o1-mini tend to perform better with concise prompts\cite{openai2024o1} - a capability Prochemy enables through its prompt refinement mechanism. 

\subsection{Time and Token Cost Analysis}
To systematically evaluate the computational efficiency of \Name{}, we analyze token consumption and processing overhead across different LLMs on three benchmarks (HumanEval, MBPP, LiveCodeBench) during inference. 
Table~\ref{tab:humaneval_mbpp} and Table~\ref{tab:livecodebench} presents the results with three metrics per configuration: mean input tokens (first value), mean output tokens (second value), and average overhead in seconds (third value), measured under controlled hardware settings.

\paragraph{Token Cost Analysis}
\Name{} achieves task-level optimization with marginal token overhead compared to baseline methods. On HumanEval with GPT-3.5-Turbo, Zero-shot + \Name{} requires $259$ input tokens and $483$ completion tokens versus $184/358$ for standard Zero-shot. This remains $15.8\%$ more efficient than AgentCoder-noiter's $304/576$ tokens. For GPT-4o, Zero-shot + \Name{} maintains better token economy than both AgentCoder and LDB. On MBPP, the total token consumption of GPT-3.5-Turbo for Zero-shot + \Name{} ($251+339=590$) shows moderate growth compared to Zero-shot ($124+191=315$) yet stays comparable to specialized methods like AgentCoder ($248+404=652$). The pattern persists in LiveCodeBench, where \Name{}'s input tokens ($589$) align with standard methods while completion tokens ($532$) remain $8.7\%$ lower than AgentCoder-noiter ($583$). Analogous phenomena are also observed in the multi-turn approach implemented within the \Name{}. 

The marginal increase in \Name{}'s input tokens arises from its optimized prompts, which incorporate richer contextual perspectives (e.g., error prevention, task decomposition in Fig.~\ref{case3}) to guide LLMs more comprehensively. Similarly, the increase in completion tokens reflects \Name{}'s focus on generating standardized, logically structured code in Fig.~\ref{case1}, with additional comments and documentation that improve code clarity and maintainability, thereby enhancing overall readability.

\paragraph{Overhead Analysis}
For time cost, \Name{} demonstrates competitive time efficiency. In GPT-3.5-Turbo HumanEval evaluations, Zero-shot + \Name{} completes in $2.79$s versus $3.76$s for standard Zero-shot - a $25.8\%$ reduction despite higher token usage. This efficiency gain becomes more pronounced with complex methods: CoT + \Name{} achieves $3.51$s latency compared to $3.91$s for vanilla CoT ($10.2\%$ improvement). The trend holds across model architectures, with Claude-3-Haiku showing $4.76$s for Zero-shot + \Name{} versus $4.95$s for AgentCoder-noiter. Multi-turn configurations reveal \Name{}'s scalability - LDB + \Name{} completes HumanEval's tasks in $11.34$s (GPT-3.5-Turbo) versus $14.49$s for standard LDB, demonstrating $21.7\%$ faster execution despite comparable token counts. The computational overhead associated with the \Name{} methodology is found to be comparable to the baseline implementation in the majority of experimental scenarios.

\paragraph{Training} 
For training, \Name{}'s token and time costs are highly sustainable compared to manual prompt engineering iterations. The optimization process incurs only a minimal one-time training cost, which does not impact inference overhead, making it especially efficient for batch processing. For instance, training on the HumanEval dataset with GPT-4o consumes approximately $18,000$ tokens in total and takes just $60$–$80$ seconds, demonstrating the low training cost. This efficiency is achieved by consolidating optimization knowledge into system prompts, rather than performing per-sample tuning, ensuring the method's practicality for real-world deployment.

% Prochemy eliminates per-sample prompt tuning by optimizing prompts holistically for entire tasks (e.g., code generation). While single-request token/time costs increase slightly, this approach reduces repetitive design and debugging overhead. On HumanEval, Zero-shot + Prochemy (517 tokens) exceeds standard Zero-shot (358 tokens) but remains below per-sample methods like AgentCoder (576 tokens). Time efficiency surpasses complex methods (e.g., AgentCoder, LDB) and even outperforms standard methods in some cases (e.g., CoT + Prochemy: 2.79s vs. CoT: 3.91s).

% For large, reusable tasks (e.g., code generation and code translation), Prochemy’s initial optimization cost is offset by long-term batch-processing efficiency.

\subsection{Threats to Validity}
\paragraph{Threats to External Validity}
The main threat to \Name{}'s external validity is that different model versions and architectures might affect its performance. To ensure our approach is generalizable, we tested it on six diverse large language models—both proprietary ones like ChatGPT and Claude, and open-source ones like DeepSeek—to validate its transferability. Although ChatGPT's o1 model recommends using simple, direct prompts~\cite{openai2024o1}, this does not conflict with \Name{}'s strategy. \Name{} identifies near-optimal prompts for each model and task using training feedback and employs a tailored, iterative prompt mutation process for each model, enabling prompt optimization that is both specific and adaptable. Our experimental results in Table~\ref{o1mini_results} further demonstrate \Name{}’s compatibility with models adhering to minimalistic prompting guidelines. 

\paragraph{Threats to Internal Validity}
A key threat to internal validity is potential data leakage, as large language models are trained on extensive open-source code that may overlap with our experimental benchmarks. However, this risk does not significantly compromise the fairness of our experiments since all methods are consistently validated under the same model configuration. Therefore, the observed relative improvements between the baseline and \Name{} are deemed reliable under these controlled conditions. To further mitigate leakage risks, we conducted additional evaluations on LiveCodeBench, a dynamically refreshed dataset updated over time to minimize historical training overlap. For the training data used or generated in our experiments, we ensure that no overlapping occurs between the training data and the test sets used to evaluate the models. This precaution helps to maintain the integrity of our experimental results and ensures that the improvements reported are due to the effectiveness of \Name{} and not from the models having prior exposure to the test data.

\section{Conclusion}
\label{sec:conclusion}
% \vspace{-1mm}

In this paper, we introduce \Name{}, an automated prompt optimization framework that systematically enhances code generation and translation across diverse large language models (LLMs). By aligning prompts with model capabilities and task requirements through iterative refinement, \Name{} eliminates manual engineering while maintaining compatibility with multi-agent and reasoning-driven workflows.
The experimental results show the effectiveness of \Name{}. On code generation benchmarks, \Name{} achieves average improvements of $4.04\%$ (zero-shot), $4.55\%$ (chain-of-thought), and $2.00\%$ (multi-turn) across four LLMs in four datasets, setting new state-of-the-art results such as $96.3\%$ accuracy on HumanEval (GPT-4o). The framework demonstrates cross-model consistency: Claude-3.5-Sonnet achieves $30.2\%$ relative improvement on LiveCodeBench’s zero-shot tasks, while DeepSeek-V3 sees $21.5\%$ gains in CoT settings. For code translation, \Name{} delivers $13.68\%$ and $8.25\%$ average improvements on Java-to-Python and Python-to-Java tasks, respectively. The further discussion shows that \Name{} also demonstrates good performance when combining with the o1-mini model, highlighting its effectiveness across different tasks and model configurations.

\Name{}’s plug-and-play design requires no architectural modifications or fine-tuning, ensuring computational efficiency. While high-capacity models exhibit diminishing returns, the framework balances automation and capability exploitation, unifying optimization across code synthesis, reasoning, and translation tasks. These results establish \Name{} as a scalable paradigm for advancing LLM-driven software development.

\section{Data Availability}
Our code, data, and results are available at \href{https://github.com/buriyuanyou/Prochemy}{https://github.com/buriyuanyou/Prochemy}.

\bibliographystyle{IEEEtran}
\bibliography{main}

\end{document}